\documentclass[default,iicol]{sn-jnl}

\usepackage{graphicx}%
\usepackage{multirow}%
\usepackage{amsmath,amssymb,amsfonts}%
\usepackage{amsthm}%
\usepackage{mathrsfs}%
\usepackage[title]{appendix}%
\usepackage[table]{xcolor}
\usepackage{textcomp}%
\usepackage{manyfoot}%
\usepackage{booktabs}%
\usepackage{algorithm}%
\usepackage{algorithmicx}%
\usepackage{algpseudocode}%
\usepackage{listings}%
\usepackage{hyperref}
\usepackage{color}
\usepackage{tabularx}
\usepackage{makecell}
\definecolor{dino}{RGB}{249,231,227}

\theoremstyle{thmstyleone}%
\theoremstyle{thmstyletwo}%

\theoremstyle{thmstylethree}%

\begin{document}

\title[iMD4GC: Incomplete Multimodal Data Integration to Advance Precise Treatment Response Prediction and Survival Analysis for Gastric Cancer]{iMD4GC: Incomplete Multimodal Data Integration to Advance Precise Treatment Response Prediction and Survival Analysis for Gastric Cancer}


\author[1]{\fnm{Fengtao} \sur{Zhou}}

\author[1]{\fnm{Yingxue} \sur{Xu}}

\author[2]{\fnm{Yanfen} \sur{Cui}}

\author[3]{\fnm{Shenyan} \sur{Zhang}}

\author[4]{\fnm{Yun} \sur{Zhu}}

\author[5]{\fnm{Weiyang} \sur{He}}

\author[6]{\fnm{Jiguang} \sur{Wang}}

\author[7]{\fnm{Xin} \sur{Wang}}

\author[8]{\fnm{Ronald} \sur{Chan}}

\author[9]{\fnm{Louis Ho Shing} \sur{Lau}}

\author[10]{\fnm{Chu} \sur{Han}}

\author[11]{\fnm{Dafu} \sur{Zhang}}

\author*[11]{\fnm{Zhenhui} \sur{Li}}

\author*[1,6]{\fnm{Hao} \sur{Chen}}

\email{jhc@cse.ust.hk}\email{lizhenhui@kmmu.edu.cn}














\affil[1]{\orgdiv{Department of Computer Science and Engineering}, \orgname{The Hong Kong University of Science and Technology}, \orgaddress{\state{Hong Kong}, \country{China}}}

\affil[2]{\orgdiv{Department of Radiology}, \orgname{Shanxi Cancer Hospital/ Shanxi Hospital Affiliated to Cancer Hospital}, \orgname{Chinese Academy of Medical Sciences/Cancer Hospital Affiliated to Shanxi Medical University},  \orgaddress{\state{Taiyuan}, \country{China}}}

\affil[3]{\orgdiv{Department of Pathology}, \orgname{The Sixth Affiliated Hospital of Sun Yat-sen University}, \orgaddress{\state{Guangzhou}, \country{China}}}

\affil[4]{\orgdiv{Department of Radiology}, \orgname{The First Affiliated Hospital of Kunming Medical University}, \orgaddress{\state{Kunming}, \country{China}}}

\affil[5]{\orgdiv{Department of Gastrointestinal Surgery}, \orgname{Sichuan Province Cancer Hospital}, \orgname{University of Electronic Science and Technology of China}, \orgaddress{\state{Chengdu}, \country{China}}}

\affil[6]{\orgdiv{Department of Chemical and Biological Engineering}, \orgname{The Hong Kong University of Science and Technology}, \orgaddress{\state{Hong Kong}, \country{China}}}

\affil[7]{\orgdiv{Department of Surgery}, \orgname{Prince of Wales Hospital}, \orgaddress{\street{The Chinese University of Hong Kong}, \state{Hong Kong}, \country{China}}}

\affil[8]{\orgdiv{Department of Anatomical and Cellular Pathology}, \orgname{Prince of Wales Hospital}, \orgaddress{\street{The Chinese University of Hong Kong}, \state{Hong Kong}, \country{China}}}

\affil[9]{\orgdiv{Department of Medicine \& Therapeutics}, \orgname{Prince of Wales Hospital}, \orgaddress{\street{The Chinese University of Hong Kong}, \state{Hong Kong}, \country{China}}}

\affil[10]{\orgdiv{Department of Radiology}, \orgname{Guangdong Provincial People's Hospital}, \orgaddress{\street{Southern Medical University}, \state{Guangzhou}, \country{China}}}

\affil[11]{\orgdiv{Department of Radiology}, \orgname{The Third Affiliated Hospital of Kunming Medical University}, \orgaddress{\street{Yunnan Cancer Hospital}, \state{Kunming}, \country{China}}}


\abstract{Gastric cancer (GC) is a prevalent malignancy worldwide, ranking as the fifth most common cancer with over 1 million new cases and 700 thousand deaths in 2020. Locally advanced gastric cancer (LAGC) accounts for approximately two-thirds of GC diagnoses, and neoadjuvant chemotherapy (NACT) has emerged as the standard treatment for LAGC. However, the effectiveness of NACT varies significantly among patients, with a considerable subset displaying treatment resistance. Ineffective NACT not only leads to adverse effects but also misses the optimal therapeutic window, resulting in lower survival rate. Hence, it is crucial to utilize clinical data to precisely predict treatment response and survival prognosis for GC patients. Existing methods relying on unimodal data falls short in capturing GC's multifaceted nature, whereas multimodal data offers a more holistic and comprehensive insight for prediction. However, existing multimodal learning methods assume the availability of all modalities for each patient, which does not align with the reality of clinical practice. The limited availability of modalities for each patient would cause information loss, adversely affecting predictive accuracy. In this study, we propose an incomplete multimodal data integration framework for GC (iMD4GC) to address the challenges posed by incomplete multimodal data, enabling precise response prediction and survival analysis. Specifically, iMD4GC incorporates unimodal attention layers for each modality to capture intra-modal information. Subsequently, the cross-modal interaction layers explore potential inter-modal interactions and capture complementary information across modalities, thereby enabling information compensation for missing modalities. To enhance the ability to handle severely incomplete multimodal data, iMD4GC employs a ``more-to-fewer'' knowledge distillation, transferring knowledge learned from more modalities to fewer ones. To evaluate iMD4GC, we collected three multimodal datasets for GC study: GastricRes (698 cases) for response prediction, GastricSur (801 cases) for survival analysis, and TCGA-STAD (400 cases) for survival analysis. The scale of our datasets is significantly larger than previous studies. The iMD4GC achieved impressive performance with an 80.2\% AUC on GastricRes, 71.4\% C-index on GastricSur, and 66.1\% C-index on TCGA-STAD, significantly surpassing other compared methods. Moreover, iMD4GC exhibits inherent interpretability, enabling transparent analysis of the decision-making process and providing valuable insights to clinicians. Furthermore, the flexible scalability provided by iMD4GC holds immense significance for clinical practice, facilitating precise oncology through artificial intelligence and multimodal data integration.}




\maketitle

\section{Introduction}
\begin{figure*}[tbp]
    \centering
    \includegraphics[width=1.0\linewidth]{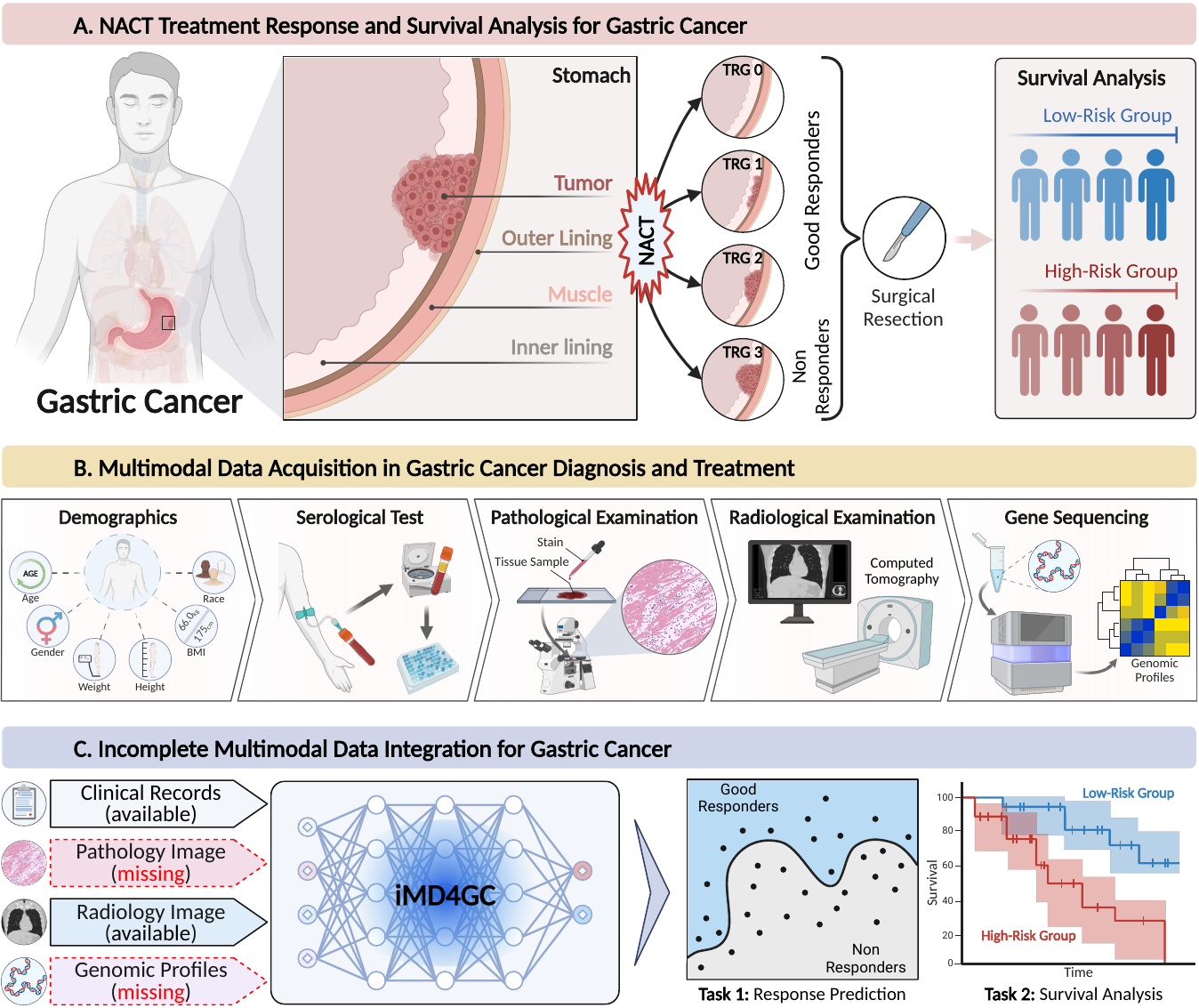}
    \caption{\textbf{Overview of treatment journey for GC patients, multimodal data acquisition in GC diagnosis and treatment, and incomplete multimodal data learning for GC.} (\textbf{A}) A schematic workflow of gastric cancer treatment and prognosis, including NACT treatment, surgical resection, and survival analysis. (\textbf{B}) The multimodal data acquisition process in the diagnosis, treatment, and prognosis of GC, involving clinical records, pathology images, radiology images, and genomic profiles. (\textbf{C}) The pipeline of incomplete multimodal data integration framework for precise response prediction and survival analysis.}
    \label{fig:data}
\end{figure*}

Gastric cancer (GC) imposes a substantial global health burden. It ranks as the fifth most prevalent cancer worldwide, standing at the fourth position among men and seventh among women~\cite{thrift2020burden}. The alarming figures from 2020 alone reveal the gravity of the situation, with over 1 million new cases diagnosed and a devastating toll of more than 700 thousand lives lost. Among GC cases, locally advanced gastric cancer (LAGC) comprises approximately two-thirds of the diagnoses~\cite{zurlo2022predictive, liu2021treatment}. To enhance treatment outcomes and prognosis for LAGC patients, neoadjuvant chemotherapy (NACT) has emerged as a promising therapeutic strategy~\cite{yoshikawa2009phase, lowy1999response}. However, the efficacy of NACT exhibits significant heterogeneity among LAGC patients, with a considerable subset displaying resistance to treatment. Research suggests that the overall response rate to NACT is less than 40\%~\cite{vauhkonen2006pathology}. Ineffective chemotherapy not only leads to adverse effects, including toxicity and financial burdens, but also deprives patients of the optimal therapeutic window. Consequently, accurate prediction of NACT response and identification of good responders (Figure~\ref{fig:data}A) are of critical importance, facilitating personalized treatment approaches and maximizing therapeutic benefits to enhance overall survival prognosis.

The multimodal data integration plays a crucial role in providing a comprehensive understanding of diseases, yielding significant benefits for the diagnosis, treatment, and prognosis of GC. The multimodal data acquired in GC diagnosis and treatment include clinical records, pathology images, radiology images, and genomic profiles, as shown in Figure~\ref{fig:data}(B). At the outset, clinicians collect essential clinical records that encompass demographic information and molecular biomarkers obtained through serological testing, which play a pivotal role in the diagnostic process. For early detection and assessment, gastrointestinal endoscopy and biopsy are used to produce pathology images (whole slide images, WSIs). These high-resolution images enable meticulous examination of tissue samples at a microscopic level, offering profound morphological insights into tumor cells and their microenvironment. Subsequently, computer tomography (CT) scans are essential for precise diagnosis and accurate tumor staging, offering insights into macroscopic features, tumor morphology, texture, and metastatic presence. Additionally, genomic profiling is pivotal in analyzing the genetics of tumor cell, shedding light on molecular changes that drive GC. By effectively integrating multimodal data in the diagnosis and treatment of GC, clinicians can achieve heightened accuracy and tailor management strategies for each patient, ultimately leading to improved outcomes. Furthermore, the analysis of multimodal data offers a comprehensive approach by amalgamating information from diverse sources, enhancing diagnostic accuracy, facilitating personalized treatment planning, and enabling the monitoring of treatment response.


In recent years, deep learning has made remarkable advancements in supporting doctors with diagnosis~\cite{wang2019rmdl, yu2021large, meng20202d, yuan2023cluster}, NACT response prediction~\cite{zhang2022deep, gu2022deep, cui2022ct}, treatment~\cite{yoshimura2020single, prendergast2020real, islam2021st}, and survival analysis~\cite{li2010survival, oh2018prediction} in GC patients. However, the majority of existing research predominantly focuses on utilizing unimodal data for prediction, which fails to capture the diverse aspects of GC and results in suboptimal performance. To surmount this limitation, multimodal learning methods have emerged to integrate complementary information from multiple modalities, enabling a more comprehensive understanding and improving prediction performance~\cite{zhang2020deep, zhong2023deep, li2023multiparametric}. Nevertheless, these multimodal learning methods generally assume the availability of all modalities for every patient, which does not align with the reality of clinical scenarios. In practice, different medical centers may adopt diverse treatment schemes and data collection protocols, leading to challenges in obtaining certain modalities (\textit{i.e.}, incomplete multimodal data). Furthermore, the high cost associated with high-throughput sequencing often renders genomic profiles unavailable for certain patients and medical centers. The incomplete multimodal data poses significant challenges for multimodal learning methods, particularly those relying on complete modalities, resulting in potential issues such as model overfitting and limited generalization. Consequently, a more robust approach is essential to address these challenges and enhance effectiveness of multimodal learning in the context of incomplete multimodal data.

This study introduces a novel multimodal learning framework, called the Incomplete Multimodal Data Integration Framework for Gastric Cancer (iMD4GC), aimed at advancing precise response prediction and survival analysis with incomplete multimodal data. Specifically, iMD4GC incorporates unimodal attention layers for each modality to capture intra-modal information. Subsequently, the cross-modal interaction layers explore potential inter-modal interactions and capture complementary information across modalities, thereby enabling information compensation for missing modalities and addressing the challenges posed by incomplete multimodal data. To further enhance its ability to handle severely incomplete multimodal data, we introduce a ``more-to-fewer'' knowledge distillation strategy, which involves distilling knowledge learned from more modalities to fewer ones, enabling improved prediction accuracy even in scenarios with more missing modalities. To evaluate the effectiveness of our proposed framework, we collect three datasets related to GC: GastricRes (698 cases) for NACT response prediction, GastricSur (801 cases) for survival analysis, and TCGA-STAD (400 cases) for survival analysis. Through extensive experiments on these datasets, we demonstrate the remarkable performance of our framework, which outperforms the compared methods by a significant margin. Moreover, our framework offers inherent interpretability, providing medical professionals and clinicians with more transparent decision-making process employed by iMD4GC. The insights gained through interpretable analysis have potential to optimize the clinical management of GC patients, leading to more personalized and effective treatment strategies. This work represents the first study on incomplete multimodal data integration for GC. Furthermore, the flexible scalability provided by iMD4GC holds immense significance for clinical practice, facilitating precise oncology through artificial intelligence and multimodal data integration.

\section{Results}
\subsection{Datasets and evaluation metrics}\label{secDATA}
To assess the effectiveness of our proposed framework in predicting NACT response and determining survival prognosis in GC patients, we collected three multimodal GC datasets from multiple hospitals: 1) GastricRes was collected from Yunnan Cancer Hospital (Kunming, China), Shanxi Cancer Hospital (Taiyuan, China), Sichuan Cancer Hospital (Chengdu, China), and the Sixth Affiliated Hospital of Sun Yat-sen University (Guangzhou, China). It comprises information from 698 patients diagnosed with gastric cancer who underwent NACT treatment. Among these cases, 325 patients exhibited a good response to the treatment, while 373 patients were classified as non-responders. 2) GastricSur was collected from the First Affiliated Hospital of Kunming Medical University (Kunming, China) and Yunnan Cancer Hospital. This dataset encompasses data from 801 patients diagnosed with gastric cancer who underwent surgical resection. 3) TCGA-STAD was sourced from The Cancer Genome Atlas (TCGA) database. It contains data from 400 patients diagnosed with gastric cancer.

The details of collected datasets are described in appendix A. Notably, these datasets suffer from significant incompleteness, with a limited number of patients possessing complete data. For instance, the GastricRes dataset contains complete data for only 240 patients, while the GastricSur dataset has complete data for 456 patients. In this study, we employed 5-fold cross-validation on these collected datasets to evaluate the performance of our framework. For the prediction of treatment response, we utilized five evaluation metrics: AUC (area under the receiver operating characteristic curve), accuracy, precision, recall, and F1-score, in which AUC is the primary metric. For survival analysis, we employed the C-index (concordance index) and time-dependent AUC~\cite{hung2010estimation, lambert2016summary} as the evaluation metrics. To comprehensively assess the performance of our framework, we calculated the mean and standard deviation for each evaluation metric across the 5-fold cross-validation.

\subsection{Compared methods}
To showcase the effectiveness and superiority of our framework, we reproduced various models for comparison, covering unimodal learning methods, multimodal learning methods, and missing modality methods. Specifically, unimodal learning methods comprise: 1) TabNet~\cite{arik2021tabnet}, a model specialized in tabular data learning. 2) SNN~\cite{klambauer2017self}, a self-normalizing neural networks. 3) Transformer~\cite{vaswani2017attention}, a classic model based on self-attention mechanisms. 4) ResNet3D~\cite{hara2018can}, a classic 3D convolutional neural network. 5) ABMIL~\cite{ilse2018attention}, an attention-based multiple instance learning (MIL) model. 6) DSMIL~\cite{li2021dual}, a dual attention-based MIL model. 7) TransMIL~\cite{shao2021transmil}, a transformer-based MIL model. 8) DTFD~\cite{zhang2022dtfd}, a double-tier feature distillation MIL model for WSI classification. 9) MHIM-MIL~\cite{tang2023multiple}, a masked hard instance mining MIL framework for WSI classification. Multimodal learning methods encompass: 1) M3IF~\cite{li2021multi}, a multi-modal multi-instance joint learning method to integrate clinical records and WSI. 2) HFBSurv~\cite{li2022hfbsurv}, a hierarchical multimodal fusion with factorized bilinear model. Missing modality methods comprise: 1) EF-LSTM~\cite{graves2013speech} and MFN~\cite{zadeh2018memory}, both of which are LSTM-based (long short-term memory) frameworks for incomplete multimodal data interactions. 2) MulT~\cite{tsai2019multimodal}, a multimodal transformer designed for unaligned multimodal language sequences and capable of handling incomplete multimodal data. 3) MMD~\cite{cui2022survival}, a multimodal deep learning framework for integrating incomplete radiology, pathology, genomic, and demographic data  in the context of brain cancer research. 4) COM~\cite{qian2023contrastive}, a contrastive masked attention model designed for incomplete multimodal learning scenarios. 5) Performer~\cite{choromanski2020rethinking} and Nystromer~\cite{xiong2021nystromformer} are variants of Transformer, known for their linear computation complexity. For the fair comparison,  we reproduced these methods under the same development environment and experimental settings as our framework.

\begin{table*}
    \begin{center}
        \fontsize{9}{11}\selectfont
        \caption{\textbf{Quantitative results for NACT response prediction.} This table presents the results of different methods on the GastricRes dataset. $\mathcal{C}$, $\mathcal{P}$, and $\mathcal{R}$ represent $\mathcal{C}$linical records,$\mathcal{R}$adiology images, and $\mathcal{P}$athology images, respectively. $+$ indicates that multiple modalities are integrated to make predictions. $*$ indicates that the dataset includes incomplete multimodal data. The best results are highlighted in \textbf{bold}, while the second-best results are \underline{underlined}.}
        \begin{tabular}{lcccccc}
            \toprule
            Method                                     & Modality                                  & AUC                            & Accuracy                       & Precision                      & Recall                         & F1-Score                       \\
            \midrule
            MLP                                        & $\mathcal{C}$linical                      & 0.665$_{\pm0.031}$             & 0.629$_{\pm0.036}$             & 0.631$_{\pm0.036}$             & 0.630$_{\pm0.037}$             & 0.626$_{\pm0.036}$             \\
            TabNet~\cite{arik2021tabnet}               & $\mathcal{C}$linical                      & 0.536$_{\pm0.020}$             & 0.512$_{\pm0.027}$             & 0.516$_{\pm0.026}$             & 0.516$_{\pm0.026}$             & 0.511$_{\pm0.027}$             \\
            Transformer~\cite{vaswani2017attention}    & $\mathcal{C}$linical                      & 0.718$_{\pm0.046}$             & 0.652$_{\pm0.035}$             & 0.678$_{\pm0.036}$             & 0.644$_{\pm0.026}$             & 0.631$_{\pm0.032}$             \\
            \midrule
            ResNet3D10~\cite{hara2018can}              & $\mathcal{R}$adiology                     & 0.648$_{\pm0.060}$             & 0.602$_{\pm0.066}$             & 0.524$_{\pm0.110}$             & 0.554$_{\pm0.044}$             & 0.521$_{\pm0.079}$             \\
            ResNet3D18~\cite{hara2018can}              & $\mathcal{R}$adiology                     & 0.651$_{\pm0.058}$             & 0.641$_{\pm0.015}$             & 0.604$_{\pm0.168}$             & 0.546$_{\pm0.054}$             & 0.478$_{\pm0.089}$             \\
            ResNet3D34~\cite{hara2018can}              & $\mathcal{R}$adiology                     & 0.669$_{\pm0.045}$             & 0.610$_{\pm0.044}$             & 0.502$_{\pm0.109}$             & 0.550$_{\pm0.051}$             & 0.506$_{\pm0.093}$             \\
            \midrule
            ABMIL~\cite{ilse2018attention}             & $\mathcal{P}$athology                     & 0.761$_{\pm0.077}$             & 0.690$_{\pm0.052}$             & 0.703$_{\pm0.064}$             & 0.648$_{\pm0.046}$             & 0.640$_{\pm0.054}$             \\
            DSMIL~\cite{li2021dual}                    & $\mathcal{P}$athology                     & 0.769$_{\pm0.078}$             & \underline{0.703$_{\pm0.043}$} & 0.696$_{\pm0.065}$             & 0.670$_{\pm0.064}$             & 0.666$_{\pm0.061}$             \\
            TransMIL~\cite{shao2021transmil}           & $\mathcal{P}$athology                     & 0.743$_{\pm0.072}$             & 0.677$_{\pm0.049}$             & 0.686$_{\pm0.058}$             & 0.692$_{\pm0.058}$             & 0.671$_{\pm0.057}$             \\
            DTFD~\cite{zhang2022dtfd}                  & $\mathcal{P}$athology                     & 0.764$_{\pm0.074}$             & 0.680$_{\pm0.066}$             & 0.692$_{\pm0.090}$             & 0.636$_{\pm0.079}$             & 0.622$_{\pm0.087}$             \\
            MHIM-MIL~\cite{tang2023multiple}           & $\mathcal{P}$athology                     & 0.760$_{\pm0.075}$             & 0.701$_{\pm0.049}$             & 0.686$_{\pm0.065}$             & 0.669$_{\pm0.063}$             & 0.669$_{\pm0.062}$             \\
            \midrule
            M3IF~\cite{li2021multi}                    & $\mathcal{C}+\mathcal{P}$                 & 0.755$_{\pm0.052}$             & 0.674$_{\pm0.083}$             & 0.664$_{\pm0.072}$             & 0.628$_{\pm0.090}$             & 0.602$_{\pm0.124}$             \\
            HFBSurv~\cite{li2022hfbsurv}               & $\mathcal{C}+\mathcal{R}+\mathcal{P}$     & 0.690$_{\pm0.087}$             & 0.529$_{\pm0.104}$             & 0.550$_{\pm0.171}$             & 0.569$_{\pm0.075}$             & 0.477$_{\pm0.134}$             \\
            EF-LSTM~\cite{graves2013speech}            & $(\mathcal{C}+\mathcal{R}+\mathcal{P})^*$ & 0.762$_{\pm0.045}$             & 0.688$_{\pm0.051}$             & 0.694$_{\pm0.049}$             & 0.685$_{\pm0.048}$             & 0.681$_{\pm0.051}$             \\
            MFN~\cite{zadeh2018memory}                 & $(\mathcal{C}+\mathcal{R}+\mathcal{P})^*$ & 0.767$_{\pm0.042}$             & 0.685$_{\pm0.052}$             & 0.687$_{\pm0.049}$             & 0.679$_{\pm0.050}$             & 0.677$_{\pm0.052}$             \\
            MulT~\cite{tsai2019multimodal}             & $(\mathcal{C}+\mathcal{R}+\mathcal{P})^*$ & 0.760$_{\pm0.047}$             & 0.659$_{\pm0.037}$             & 0.676$_{\pm0.044}$             & 0.651$_{\pm0.035}$             & 0.642$_{\pm0.036}$             \\
            MMD~\cite{cui2022survival}                 & $(\mathcal{C}+\mathcal{R}+\mathcal{P})^*$ & 0.741$_{\pm0.059}$             & 0.672$_{\pm0.060}$             & 0.689$_{\pm0.057}$             & 0.672$_{\pm0.063}$             & 0.661$_{\pm0.064}$             \\
            COM~\cite{qian2023contrastive}             & $(\mathcal{C}+\mathcal{R}+\mathcal{P})^*$ & 0.760$_{\pm0.051}$             & 0.646$_{\pm0.094}$             & 0.546$_{\pm0.227}$             & 0.627$_{\pm0.105}$             & 0.565$_{\pm0.177}$             \\
            Performer~\cite{choromanski2020rethinking} & $(\mathcal{C}+\mathcal{R}+\mathcal{P})^*$ & 0.748$_{\pm0.049}$             & 0.682$_{\pm0.041}$             & 0.693$_{\pm0.041}$             & 0.681$_{\pm0.039}$             & 0.674$_{\pm0.042}$             \\
            Nystromer~\cite{xiong2021nystromformer}    & $(\mathcal{C}+\mathcal{R}+\mathcal{P})^*$ & \underline{0.774$_{\pm0.040}$} & \underline{0.703$_{\pm0.047}$} & \underline{0.711$_{\pm0.040}$} & \underline{0.705$_{\pm0.046}$} & \underline{0.699$_{\pm0.048}$} \\
            \midrule
            iMD4GC                                     & $(\mathcal{C}+\mathcal{R}+\mathcal{P})^*$ & \textbf{0.802$_{\pm0.050}$}    & \textbf{0.758$_{\pm0.043}$}    & \textbf{0.764$_{\pm0.062}$}    & \textbf{ 0.715$_{\pm0.087}$}   & \textbf{0.753$_{\pm0.027}$}    \\
            \bottomrule
        \end{tabular}
    \end{center}
    \label{tab:response}
\end{table*}

\begin{table*}
    \begin{center}
        \fontsize{9}{11}\selectfont
        \caption{\textbf{Quantitative results for survival prediction.} This table presents the results of different methods on the GastricSur and TCGA-STAD datasets. $\mathcal{C}$, $\mathcal{P}$, $\mathcal{R}$, and $\mathcal{G}$ represent $\mathcal{C}$linical records, $\mathcal{P}$athology images, $\mathcal{R}$adiology images, and $\mathcal{G}$enomic profiles, respectively. $+$ indicates that multiple modalities are integrated to make predictions. $*$ indicates that the dataset includes incomplete multimodal data. The best results are highlighted in \textbf{bold}, while the second-best results are \underline{underlined}.}
        \begin{tabular}{l|ccc|ccc}
            \toprule
            {\multirow{2}{*}{Method}}                  & \multicolumn{3}{c|}{GastricSur}           & \multicolumn{3}{c}{TCGA-STAD}                                                                                                                                                 \\\cmidrule(lr){2-4}\cmidrule(lr){5-7}
                                                       & Modality                                  & C-Index                        & AUC                            & Modality                                  & C-Index                        & AUC                            \\
            \midrule
            MLP                                        & $\mathcal{C}$linical                      & 0.622$_{\pm0.016}$             & 0.649$_{\pm0.054}$             & $\mathcal{C}$linical                      & 0.499$_{\pm0.024}$             & 0.491$_{\pm0.032}$             \\
            TabNet~\cite{arik2021tabnet}               & $\mathcal{C}$linical                      & 0.540$_{\pm0.029}$             & 0.584$_{\pm0.037}$             & $\mathcal{C}$linical                      & 0.583$_{\pm0.021}$             & 0.586$_{\pm0.033}$             \\
            Transformer~\cite{vaswani2017attention}    & $\mathcal{C}$linical                      & 0.600$_{\pm0.038}$             & 0.628$_{\pm0.039}$             & $\mathcal{C}$linical                      & 0.548$_{\pm0.034}$             & 0.566$_{\pm0.054}$             \\
            \midrule
            ResNet3D10~\cite{hara2018can}              & $\mathcal{R}$adiology                     & 0.623$_{\pm0.040}$             & 0.642$_{\pm0.056}$             & $\mathcal{R}$adiology                     & -                              & -                              \\
            ResNet3D18~\cite{hara2018can}              & $\mathcal{R}$adiology                     & 0.628$_{\pm0.035}$             & 0.646$_{\pm0.036}$             & $\mathcal{R}$adiology                     & -                              & -                              \\
            ResNet3D34~\cite{hara2018can}              & $\mathcal{R}$adiology                     & 0.583$_{\pm0.057}$             & 0.611$_{\pm0.078}$             & $\mathcal{R}$adiology                     & -                              & -                              \\
            \midrule
            MLP                                        & $\mathcal{G}$enomics                      & -                              & -                              & $\mathcal{G}$enomics                      & 0.572$_{\pm0.052}$             & 0.582$_{\pm0.056}$             \\
            SNN~\cite{klambauer2017self}               & $\mathcal{G}$enomics                      & -                              & -                              & $\mathcal{G}$enomics                      & 0.602$_{\pm0.038}$             & 0.610$_{\pm0.033}$             \\
            Nystromer~\cite{xiong2021nystromformer}    & $\mathcal{G}$enomics                      & -                              & -                              & $\mathcal{G}$enomics                      & 0.584$_{\pm0.052}$             & 0.606$_{\pm0.052}$             \\
            \midrule
            ABMIL~\cite{ilse2018attention}             & $\mathcal{P}$athology                     & 0.642$_{\pm0.015}$             & 0.644$_{\pm0.032}$             & $\mathcal{P}$athology                     & 0.620$_{\pm0.030}$             & 0.654$_{\pm0.048}$             \\
            DSMIL~\cite{li2021dual}                    & $\mathcal{P}$athology                     & 0.648$_{\pm0.018}$             & 0.639$_{\pm0.026}$             & $\mathcal{P}$athology                     & 0.630$_{\pm0.021}$             & 0.662$_{\pm0.033}$             \\
            TransMIL~\cite{shao2021transmil}           & $\mathcal{P}$athology                     & 0.648$_{\pm0.030}$             & 0.665$_{\pm0.036}$             & $\mathcal{P}$athology                     & 0.617$_{\pm0.054}$             & 0.636$_{\pm0.075}$             \\
            DTFD~\cite{zhang2022dtfd}                  & $\mathcal{P}$athology                     & 0.641$_{\pm0.015}$             & 0.636$_{\pm0.027}$             & $\mathcal{P}$athology                     & 0.605$_{\pm0.021}$             & 0.609$_{\pm0.045}$             \\
            MHIM-MIL~\cite{tang2023multiple}           & $\mathcal{P}$athology                     & 0.639$_{\pm0.017}$             & 0.618$_{\pm0.042}$             & $\mathcal{P}$athology                     & 0.615$_{\pm0.026}$             & 0.651$_{\pm0.034}$             \\
            \midrule
            M3IF~\cite{li2021multi}                    & $\mathcal{C}+\mathcal{P}$                 & \underline{0.674$_{\pm0.018}$} & \underline{0.689$_{\pm0.028}$} & $\mathcal{C}+\mathcal{P}$                 & \underline{0.649$_{\pm0.036}$} & \underline{0.670$_{\pm0.062}$} \\
            HFBSurv~\cite{li2022hfbsurv}               & $\mathcal{C}+\mathcal{R}+\mathcal{P}$     & 0.656$_{\pm0.049}$             & 0.685$_{\pm0.071}$             & $\mathcal{C}+\mathcal{P}+\mathcal{G}$     & 0.599$_{\pm0.054}$             & 0.633$_{\pm0.066}$             \\
            EF-LSTM~\cite{graves2013speech}            & $(\mathcal{C}+\mathcal{R}+\mathcal{P})^*$ & 0.615$_{\pm0.037}$             & 0.620$_{\pm0.056}$             & $(\mathcal{C}+\mathcal{P}+\mathcal{G})^*$ & 0.626$_{\pm0.037}$             & 0.660$_{\pm0.053}$             \\
            MFN~\cite{zadeh2018memory}                 & $(\mathcal{C}+\mathcal{R}+\mathcal{P})^*$ & 0.615$_{\pm0.028}$             & 0.611$_{\pm0.032}$             & $(\mathcal{C}+\mathcal{P}+\mathcal{G})^*$ & 0.631$_{\pm0.037}$             & 0.665$_{\pm0.048}$             \\
            MulT~\cite{tsai2019multimodal}             & $(\mathcal{C}+\mathcal{R}+\mathcal{P})^*$ & 0.579$_{\pm0.045}$             & 0.565$_{\pm0.045}$             & $(\mathcal{C}+\mathcal{P}+\mathcal{G})^*$ & 0.600$_{\pm0.056}$             & 0.629$_{\pm0.062}$             \\
            MMD~\cite{cui2022survival}                 & $(\mathcal{C}+\mathcal{R}+\mathcal{P})^*$ & 0.607$_{\pm0.042}$             & 0.634$_{\pm0.056}$             & $(\mathcal{C}+\mathcal{P}+\mathcal{G})^*$ & 0.592$_{\pm0.039}$             & 0.608$_{\pm0.059}$             \\
            COM~\cite{qian2023contrastive}             & $(\mathcal{C}+\mathcal{R}+\mathcal{P})^*$ & 0.552$_{\pm0.031}$             & 0.561$_{\pm0.033}$             & $(\mathcal{C}+\mathcal{P}+\mathcal{G})^*$ & 0.625$_{\pm0.017}$             & 0.665$_{\pm0.021}$             \\
            Performer~\cite{choromanski2020rethinking} & $(\mathcal{C}+\mathcal{R}+\mathcal{P})^*$ & 0.622$_{\pm0.027}$             & 0.612$_{\pm0.032}$             & $(\mathcal{C}+\mathcal{P}+\mathcal{G})^*$ & 0.618$_{\pm0.038}$             & 0.647$_{\pm0.053}$             \\
            Nystromer~\cite{xiong2021nystromformer}    & $(\mathcal{C}+\mathcal{R}+\mathcal{P})^*$ & 0.651$_{\pm0.026}$             & 0.662$_{\pm0.045}$             & $(\mathcal{C}+\mathcal{P}+\mathcal{G})^*$ & 0.637$_{\pm0.054}$             & 0.650$_{\pm0.074}$             \\
            \midrule
            iMD4GC                                     & $(\mathcal{C}+\mathcal{R}+\mathcal{P})^*$ & \textbf{0.714$_{\pm0.008}$}    & \textbf{0.738$_{\pm0.019}$}    & $(\mathcal{C}+\mathcal{P}+\mathcal{G})^*$ & \textbf{0.661$_{\pm0.058}$}    & \textbf{0.690$_{\pm0.072}$}    \\
            \bottomrule
        \end{tabular}
    \end{center}
    \label{tab:survival}
\end{table*}

\subsection{iMD4GC accurately identifies good responders and non-responders}
The quantitative results, as depicted in Table~\ref{tab:response}, provide a comprehensive evaluation of different methods applied to the GastricRes dataset for predicting NACT response. Notably, pathology images outperform clinical records and radiology images in predicting NACT response, demonstrating their superior informativeness. All pathology-centric methods consistently outperform the other two modalities, underscoring the efficacy of pathology images in predicting treatment response. However, it is essential to note that integrating pathology images with other modalities resulted in a performance degradation. For example, M3IF~\cite{li2021multi} attained 75.5\% AUC, 67.4\% ACC, 66.4\% Precision, 62.8\% Recall, and 60.2\% F1-Score, which were 0.6\%, 1.6\%, 3.9\%, 2.0\%, and 3.8\% lower than ABMIL~\cite{ilse2018attention}, respectively. This decline in performance could be attributed to the inherent heterogeneity among different modalities, resulting in the challenge of multimodal feature fusion. Furthermore, the introduction of additional modalities led to further performance degradation. For example, HFBSurv~\cite{li2022hfbsurv} only achieved 69.0\% AUC, 52.9\% ACC, 55.0\% Precision, 56.9\% Recall, and 47.7\% F1-Score. On one hand, incorporating more modalities could exacerbate the dilemma of heterogeneous data fusion. On the other hand,  these multimodal learning methods require the availability of all modalities. More involved modalities would lead to more discarded data due to the actual modality incompleteness, resulting in under-fitting of the models.

\begin{figure*}[tbp]
    \centering
    \includegraphics[width=1.0\linewidth]{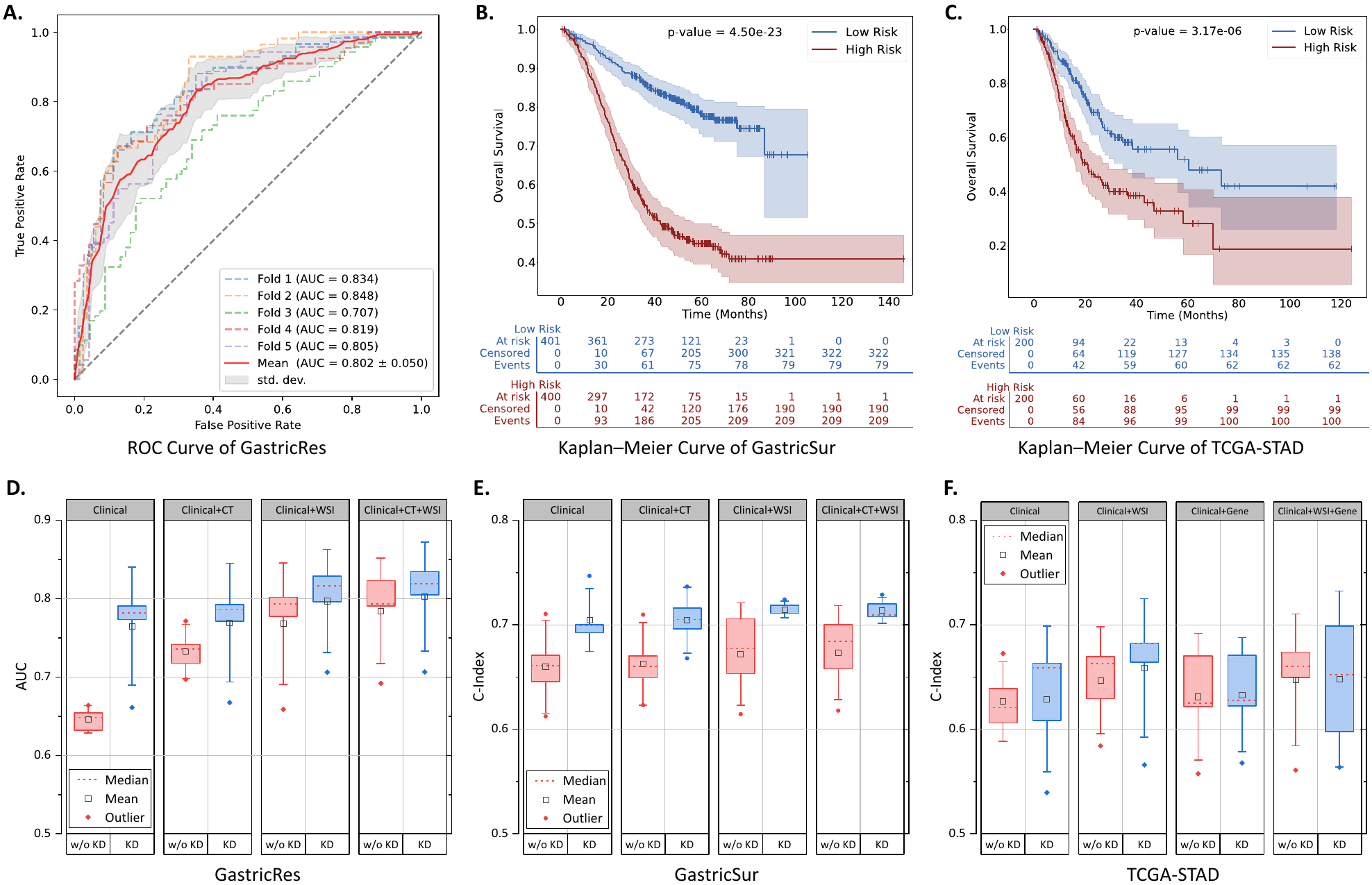}
    \caption{\textbf{ROC curves for response prediction, significance analysis for survival prediction, and comparative analysis for knowledge distillaiton.} (\textbf{A}) ROC curve and AUC value of each fold on GastricRes dataset. The AUC values consistently exceed 80.0\% across most folds, except for the third fold. (\textbf{B-C}) Kaplan-Meier curves on GastricSur and TCGA-STAD datasets: all patients are stratified into \textcolor{blue}{low-risk group} and \textcolor{red}{high-risk group} according to the predicted risk scores. The tables provide additional information about the number of individuals at risk at each time point. (\textbf{D-F}) Performance comparison on three datasets \textcolor{red}{before} and \textcolor{blue}{after} knowledge distillation.}
    \label{fig:performance}
\end{figure*}

Incomplete multimodal data would cause valuable information loss and disrupt the synergistic effects between modalities, leading to undesirable performance degradation. For instance, MMD~\cite{cui2022survival} achieved 74.1\% AUC, and Performer~\cite{choromanski2020rethinking} achieved 74.8\% AUC on the GastricRes dataset, both of which were 1.4\% and 0.7\% lower than M3IF, respectively. Despite the existence of dedicated methods for handling incomplete multimodal data, such as MulT~\cite{tsai2019multimodal} and COM~\cite{qian2023contrastive}, their performance still fell short of significantly outperforming pathology-centric methods. In contrast, our proposed framework consistently outperformed the compared methods by a substantial margin across multiple evaluation metrics. Specifically, our framework achieved impressive results, with 80.2\% AUC, 75.8\% ACC, 76.4\% Precision, 71.5\% Recall, and 75.3\% F1-Score, surpassing other methods in terms of predictive accuracy and reliability. When compared with the second-best method, Nystromer~\cite{xiong2021nystromformer}, our framework exhibited superiority by 2.8\%, 5.5\%, 5.3\%, 1.0\%, 5.4\% in terms of AUC, ACC, Precision, Recall, and F1-Score, respectively. These results substantiated the effectiveness of our proposed framework in predicting the NACT response, particularly when dealing with incomplete multimodal data. In addition, we presented the ROC curve for each fold in Figure~\ref{fig:performance}(B), where it can be observed that the AUC values consistently exceed 80.0\% across most folds, except for the third fold. Further investigation revealed that the number of pathology images in the third fold is significantly lower than that in the other folds. This observation further underscores the importance of pathology images in accurately predicting the NACT response.

\subsection{iMD4GC accurately stratifies low-risk and high-risk patients}
The quantitative results regarding survival prediction, as presented in Table~\ref{tab:survival}, encompass the GastricSur and TCGA-STAD datasets. Similar to the observations made in NACT response prediction, almost all pathology-centric methods surpassed the models using clinical records, radiology images, and even genomic profiles in terms of survival prediction. It is noteworthy that the performance of radiology images was slightly superior to that of clinical records. This could be attributed to the macroscopic features, morphological characteristics, tumor texture, and the presence of metastasis depicted in radiology images, which serve as significant indicators for survival prediction. These visual features can be effectively detected and captured by deep learning models, thereby mitigating the impact of inter-modal heterogeneity when conducting multimodal feature fusion. For instance, HFBSurv~\cite{li2022hfbsurv} achieved 69.2\% C-index and 68.5\% AUC on GastricSur dataset, which were 1.4\% and 4.1\% higher than ABMIL~\cite{ilse2018attention}, respectively. However, the performance of HFBSurv~\cite{li2022hfbsurv} on TCGA-STAD dataset was significantly lower than that of ABMIL, with 59.9\% C-index and 63.3\% AUC. That means, the heterogeneity between pathology images and genomic profiles still hinders the effective fusion of these two modalities. As for incomplete multimodal data scenario, the performance of most missing modality methods was still inferior to that of the pathology-centric methods.

Specifically, our framework achieved 71.4\% C-index and 73.8\% AUC on GastricSur, which were 4.0\% and 4.9\% higher than the second-best method, M3IF~\cite{li2021multi}, respectively. On TCGA-STAD, our framework attained 66.1\% C-index and 69.0\% AUC, which were 1.2\% and 2.0\% higher than M3IF, respectively. These results substantiated the effectiveness of our proposed framework in survival prediction when utilizing incomplete multimodal data. To further validate the effectiveness of our proposed framework, we stratified all patients into low-risk and high-risk groups based on the mid-value of the predicted risk scores. Subsequently, we employed Kaplan-Meier curves to visualize the survival events of all patients. The analysis results were presented in Figure~\ref{fig:performance}(B-C). Additionally, we utilized the Logrank test ($p$-value) to measure the statistical significance between the low-risk group (blue) and the high-risk group (red). As observed in this figure, the $p$-values associated with our framework ($4.50\times 10^{-23}$ for the GastricSur and $3.17\times 10^{-6}$ for the TCGA-STAD) were significantly lower than 0.05. These findings further reinforced the effectiveness of our proposed framework in survival analysis.

\begin{figure*}[tbp]
    \centering
    \includegraphics[width=1.0\linewidth]{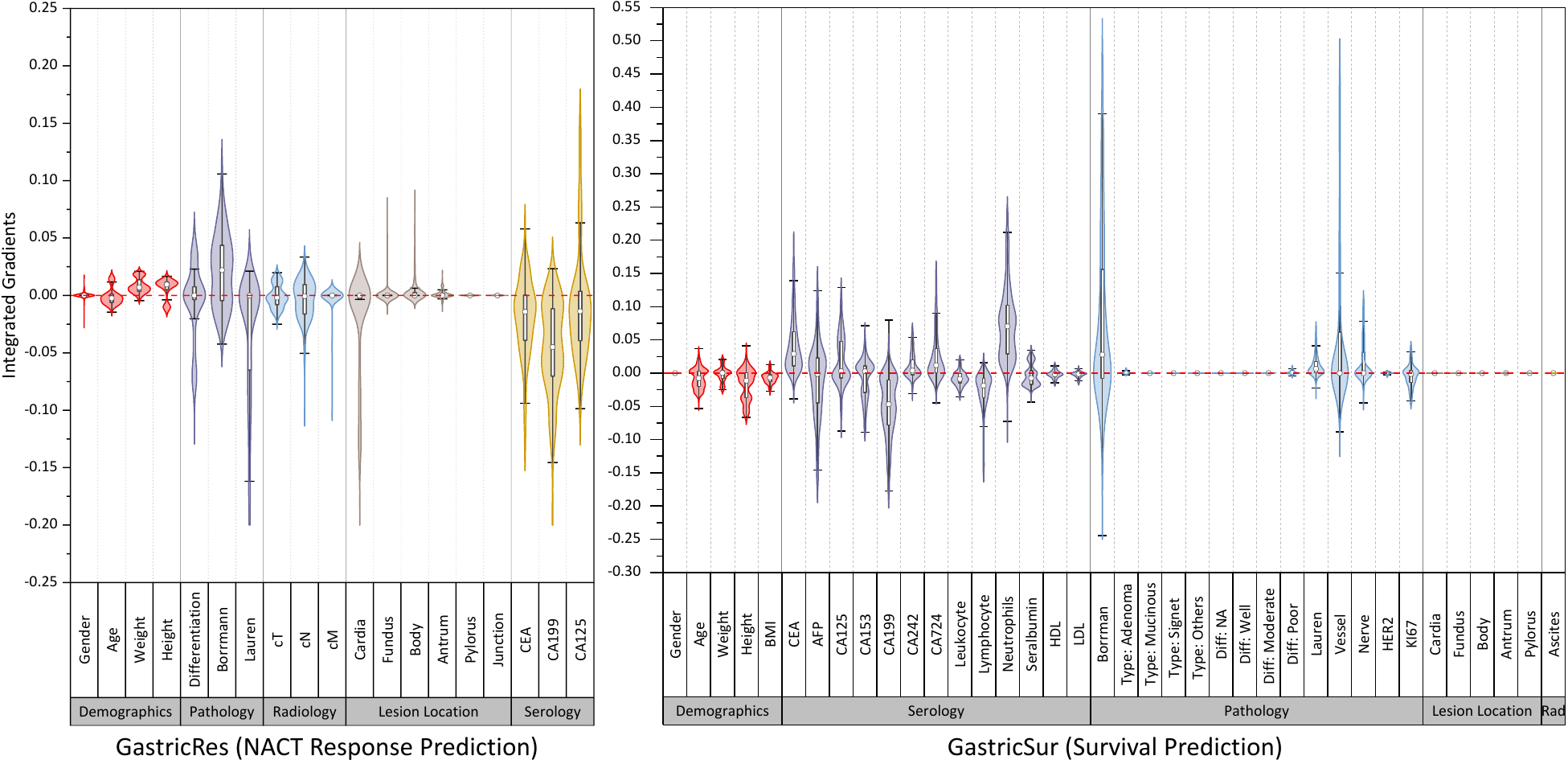}
    \caption{\textbf{Contribution of individual clinical record for NACT response prediction and survival analysis.} There are 14 clinical records involved in the GastricRes dataset and 28 clinical records involved in the TCGA-STAD dataset. The contribution values are calculated by Integrated Gradients~\cite{sundararajan2017axiomatic}. Positive contribution values signify a positive influence on the model prediction (response prediction and death risk), while negative values indicate a negative influence. Conversely, zero contribution values imply that the corresponding records have negligible impact on the model prediction.}
    \label{fig:clinical}
\end{figure*}

\subsection{Knowledge distillation unlocks the performance of iMD4GC}
In general, the availability of more modalities leads to a more comprehensive information compared to fewer modalities. As mentioned in section~\ref{secDATA}, the datasets collected for this study suffer from significant incompleteness. To further enhance the performance of our framework when dealing with severely incomplete multimodal data, we propose a ``more-to-fewer'' knowledge distillation, aiming to transfer the knowledge acquired from all available modalities to the reduced subset. To evaluate the effectiveness of this approach, we conducted comparative experiments on three datasets. It is worth noting that clinical records serve as the fundamental information in clinical practice and are available for all patients. In our comparative analysis, we considered clinical records as the baseline and evaluated the performance of our framework on different subsets of modalities. The experimental results on three datasets are presented in Figure~\ref{fig:performance}(D-F).

When considering only clinical records, our framework achieved 64.6\% AUC on GastricRes, 66.0\% C-index on GastricSur, and 62.7\% C-index on TCGA-STAD, respectively. After applying the knowledge distillation, the performance of our framework on clinical records exhibited significant improvements, achieving 76.5\% AUC on GastricRes, 70.2\% C-index on GastricSur, and 62.9\% C-index on TCGA-STAD, respectively. As the number of available modalities increased, the performance of our framework gradually improved. Although the performance on datasets with fewer available modalities still lagged behind those with more available modalities, the performance gap was substantially reduced. It is important to note that the benefits derived from knowledge distillation are not universally applicable in all scenarios. For instance, in the TCGA-STAD dataset, clinical records alone may lack the discriminative information required for accurate survival prediction, rendering knowledge distillation ineffective in improving the performance of our framework solely based on clinical records. In such cases, the inclusion of other informative modalities becomes necessary to enhance performance.

\begin{figure*}[tbp]
    \centering
    \includegraphics[width=0.95\linewidth]{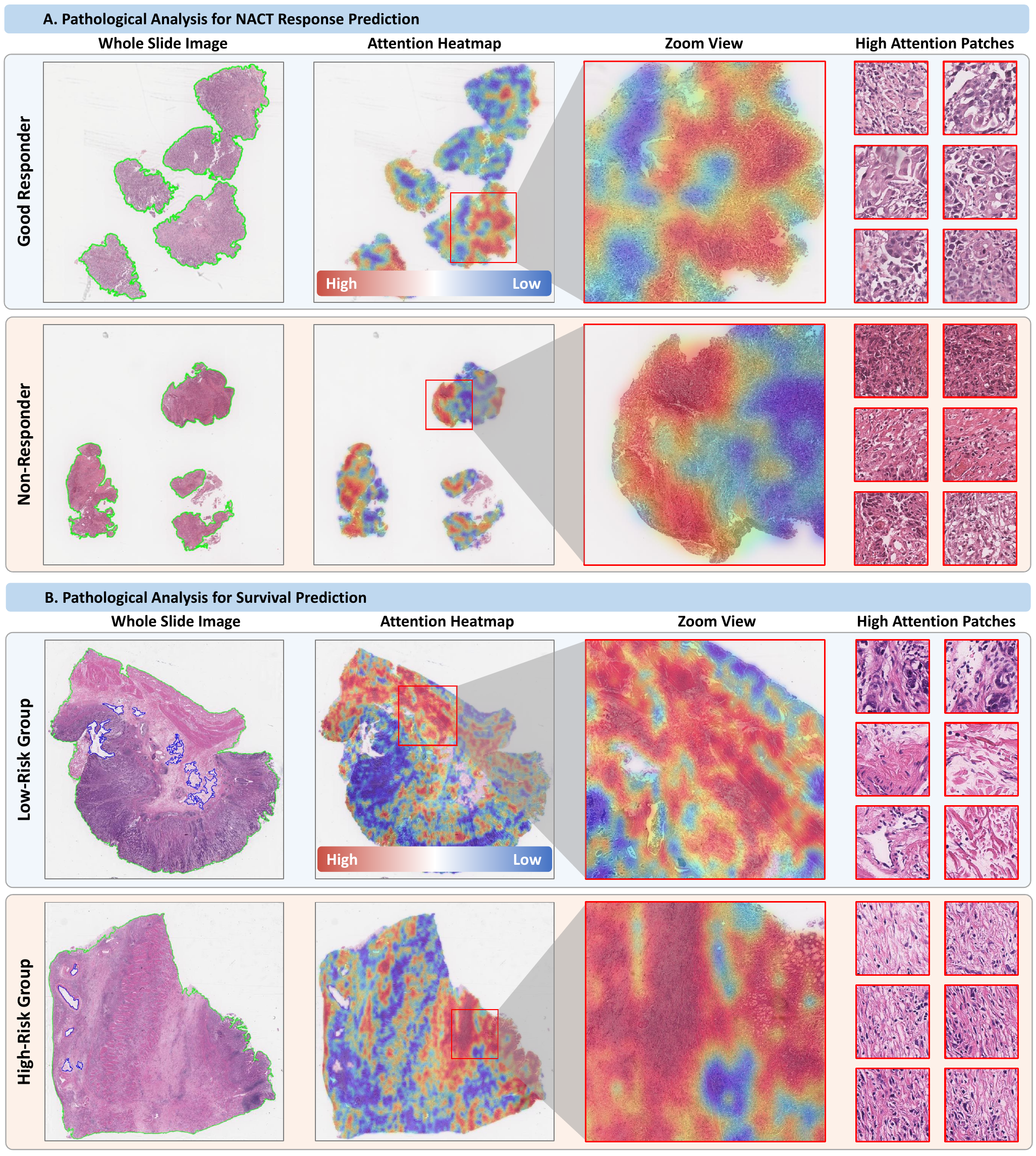}
    \caption{\textbf{Heatmaps for pathological analysis.} (\textbf{A}) Visualization comparison between good responders and non-responders. (\textbf{B}) Visualization comparison between low-risk and high-risk patients. The regions with high attention scores are deemed more valuable for model prediction, while those with low attention scores carry less significance. Right column shows the top-6 patches with the highest attention scores.}
    \label{fig:pathology}
\end{figure*}

\subsection{iMD4GC reveals contribution of individual clinical record}
Clinical records represent valuable sources of tabular data that provide essential information about the medical history of patient, including demographic factors, serological testing, and pathological examination, among others. Prior studies and experiments have demonstrated the efficacy of clinical records in predicting NACT response and survival prognosis. However, the specific contribution of individual clinical record to the model predictions remains uncertain. To gain insights into the value of different clinical records, we employ Integrated Gradients~\cite{sundararajan2017axiomatic} to calculate the contribution of each clinical record in the model decision-making process, as illustrated in Figure~\ref{fig:clinical}. These contributions provide crucial insights into the direction and magnitude of influence. Positive contributions indicate a positive influence on the model prediction, while negative values suggest a negative influence. Conversely, zero contribution values imply that the corresponding records have a negligible impact on the model prediction. To mitigate randomness, we focus on clinical records with significant absolute attribution values, taking into account both the mean and median values. The analysis results are presented in Figure~\ref{fig:clinical}. This figure illustrates that lesion location and tumor stage have minimal impact on the model decision-making process. Similarly, demographic factors like gender, age, and BMI have little influence. In contrast, records associated with serological testing and pathological examination hold greater importance. These findings underscore the effectiveness of our framework in identifying key clinical records for predicting NACT response and survival prognosis, offering critical insights for clinical decision-making.

\subsection{iMD4GC enables identification of pathological patterns}
In the field of computational pathology, the analysis of WSIs poses a challenge due to their substantial size, making it difficult to directly input them into deep learning models. To address this issue, it becomes necessary to divide the WSIs into smaller patches and input them into the model for prediction. However, not all patches have equal importance in predicting NACT response and survival prognosis. In our iMD4GC framework, we utilized attention scores to identify the discriminative patterns and reveal the relative importance of each pathological patch in the model predictions, as depicted in Figure~\ref{fig:pathology}(A-B). Patches with high attention scores are considered more valuable for the model predictions, while those with low attention scores carry less significance. Notably, our proposed framework demonstrates the capability to identify morphology specific to different prediction tasks, even in the absence of pixel-level and patch-level annotations. This indicates that our model can learn meaningful features directly from the WSIs, enabling accurate prediction without the need for detailed annotations. By leveraging attention scores, iMD4GC provides insights into the pathological patterns that contribute significantly to our predictive models, enhancing our understanding of the underlying factors influencing NACT response and survival prognosis.

\begin{figure*}[tbp]
    \centering
    \includegraphics[width=1.0\linewidth]{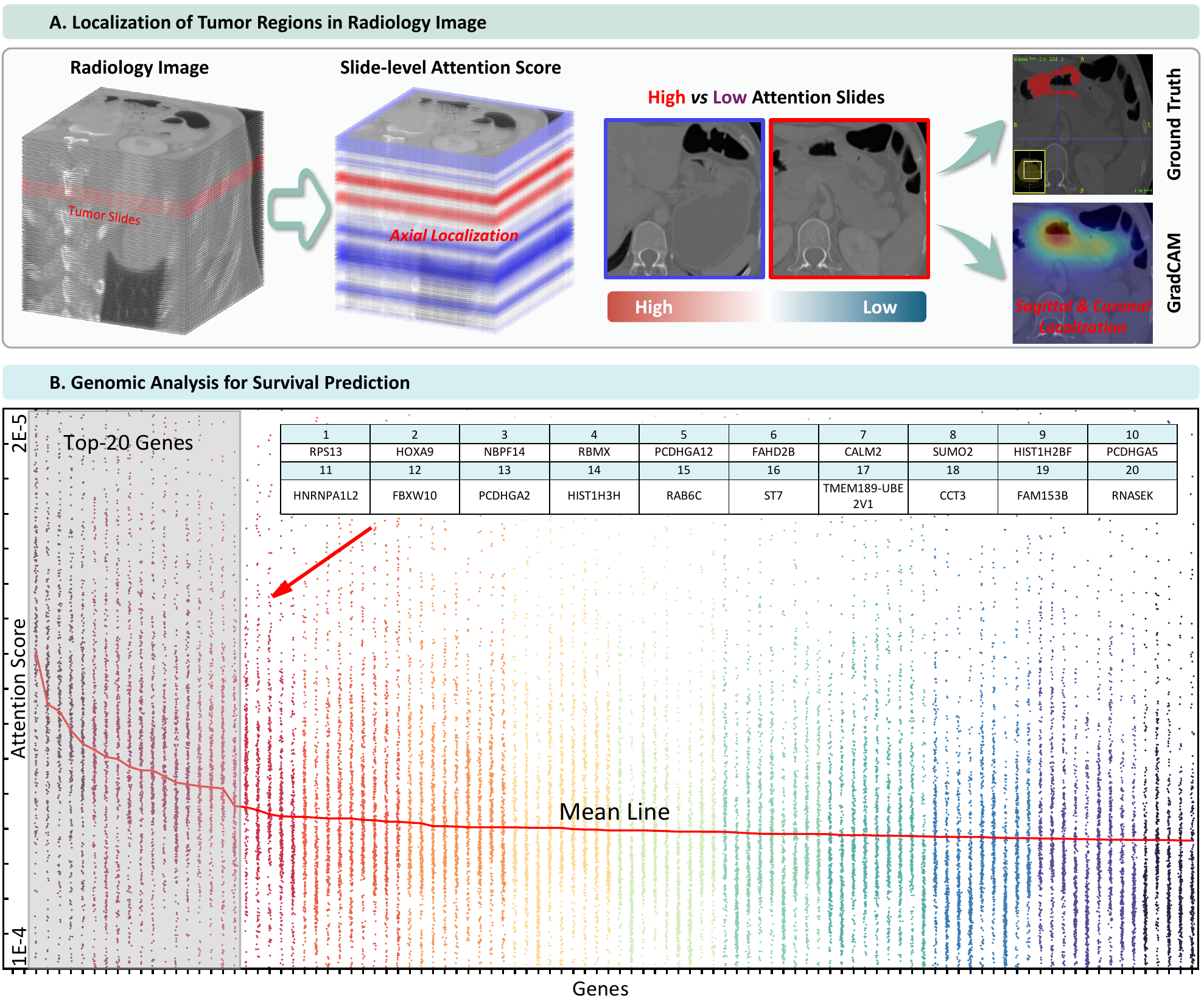}
    \caption{\textbf{Radiological analysis and genetic analysis.} (\textbf{A}) An example of localization results for radiology images. The attention scores in iMD4GC provides the axial localization of the tumor region first. And then, utilizing Grad-CAM shows the sagittal and coronal localization of the tumor region. (\textbf{B}) The attention distribution of the top-100 genes for survival prediction. Each point represents the attention score of one sample. The table shows the top-20 genes with the highest contribution values.}
    \label{fig:radiology}
\end{figure*}

\subsection{iMD4GC enables localization of tumor region in radiology images}
In previous studies focusing on NACT response prediction and survival analysis for GC, many of them heavily relied on radiomics features extracted from radiology images, which necessitate manual delineation of the tumor region at the pixel level by radiologists. However, this process is extremely time-consuming and labor-intensive. In our framework, we circumvent the need for pixel-level annotations by directly inputting radiology images into the model for prediction. Nevertheless, our framework still provides a coarse localization of the tumor region. Taking inspiration from interpretability techniques employed in pathology, we have incorporated attention scores to reveal the relative importance of each slice in the radiology images for model predictions, thereby offering axial localization of the tumor region. Subsequently, we utilized Grad-CAM~\cite{selvaraju2017grad} to highlight the relative importance of pixels in each slice for the model predictions, enabling sagittal and coronal localization of the tumor region. There is a localization example depicted in Figure~\ref{fig:radiology}(A). When compared to the ground truth annotated by radiologists, the localization results exhibit a favorable level of accuracy, they can provide valuable information to aid radiologists in locating the tumor and delineating its region. The consistency observed between the localization results and the ground truth further validates the effectiveness of our proposed framework in predicting NACT response and survival prognosis.

\subsection{iMD4GC enables potential biomarker discovery in genomics}
In this study, we utilized RNA-seq as the genomic profile, which enables the measurement of gene expression levels. This information offers valuable insights into the functional activity of genes and their involvement in various biological processes. However, genomic profiles are characterized by their exceptionally high dimensionality, necessitating an analysis of the relative importance of each gene in the model predictions. To address this, we utilize attention scores to unveil the relative importance of each gene in the model predictions. In Figure~\ref{fig:radiology}(B), we present the distribution of attention scores for the top 100 genes based on their contribution values. Additionally, we provide a list of the top 20 genes that demonstrate potential associations with survival prediction in gastric cancer. These findings highlight the significance of specific genes in influencing the model predictions and offer potential targets for further investigation and understanding of gastric cancer prognosis.

Among these listed genes, HOXA9, a transcription factor renowned for its role in embryonic development and cellular differentiation, has been linked to gastric cancer progression~\cite{yang2009tumorigenic, ma2017high, shenoy2023hoxa9}. Its upregulation suggests a potential connection to unfavorable survival outcomes in gastric cancer patients. RBMX, an RNA-binding protein, displays altered expression in gastric cancer~\cite{ge2018proteomic, yan2021rbmx, wang2022construction}, potentially contributing to tumorigenesis and impacting patient survival. CALM2, a calcium-binding protein involved in intracellular signaling, has been implicated in various cancers, including gastric cancer~\cite{mu2021calmodulin, sun2023knockdown}. Dysregulation of CALM2 could influence survival outcomes in gastric cancer patients. Additionally, SUMO2, a protein modifier involved in post-translational modifications, has been associated with gastric cancer development and progression~\cite{hu2021nsun2}. Further exploration is required to comprehend the functions, interactions, and prognostic significance of the remaining genes in survival prediction for gastric cancer.

\section{Discussion}
As one of the most prevalent malignancies worldwide, GC has attracted increasing attention in the field of artificial intelligence (AI) research. However, current AI research in GC primarily focuses on unimodal data applications, limiting the comprehensive understanding of this disease. Although some studies have proposed multimodal learning methods to integrate complementary information from multiple modalities, these approaches often assume the availability of all modalities for each patient, which does not align with clinical reality. To tackle the challenges posed by incomplete multimodal data, we present the first work on incomplete multimodal data integration for GC (iMD4GC), enabling precise prediction of treatment response and survival analysis with incomplete multimodal data. Through extensive experiments on three collected datasets, we demonstrate that our proposed iMD4GC achieves promising performance in terms of response prediction and survival analysis, outperforming other compared methods by a substantial margin.

In addition to impressive prediction performance in predicting NACT treatment response and survival prognosis for GC patients, iMD4GC goes beyond by offering inherent interpretability, which enables in-depth analysis of the decision-making process. This inherent interpretability represents a significant breakthrough in addressing concerns associated with the black-box nature of deep learning models, which often hinder their trustworthiness, transparency, and accountability in clinical applications. In this study, we conducted a comprehensive and rigorous interpretable analysis of iMD4GC, aiming to shed light on the distinct contributions of each modality in predicting treatment response and conducting survival analysis. Notably, this analysis was carried out in close collaboration with a team of experts, including clinicians, radiologists, pathologists, and biologists, ensuring the validity and clinical relevance of our findings. Through this collaborative effort, we validated that the predictions made by iMD4GC are firmly grounded in reasonable clinical assumptions, bolstering the confidence of clinicians in the model outputs and enabling them to make informed decisions in patient care.

Furthermore, the interpretability of iMD4GC has been proved to be instrumental in identifying discriminative features and patterns within the multimodal data that significantly influence the prediction of treatment response and survival prognosis. For instance, we employed Integrated Gradients~\cite{sundararajan2017axiomatic} to elucidate the contribution of individual clinical record and leveraged attention scores to highlight the pivotal genes within the genomic profiles. Importantly, these findings possess the potential to catalyze the discovery of new biomarkers or therapeutic targets, thereby improving GC treatment and management. The ability of iMD4GC to uncover such clinically relevant information further underscores its significance in advancing our understanding of GC and ultimately improving patient outcomes.

The iMD4GC stands out by providing not only inherent interpretability but also flexible scalability for integrating multimodal data. In the rapidly evolving era of information technology, the acquisition and collection of multimodal data across various medical institutions have become increasingly accessible. This accessibility opens up exciting possibilities for leveraging multiple data sources to gain a comprehensive understanding of diseases, thereby enhancing clinical decision-making and patient care. Consequently, researchers have devoted significant efforts to developing multimodal learning methods that explore potential correlations among different modalities and leverage their integration for predictive purposes. However, it is important to note that most existing multimodal learning methods have inherent limitations. They typically focus on predetermined modalities, restricting their adaptability in real-world scenarios. These methods lack the flexibility to incorporate additional multimodal data, preventing researchers and clinicians from fully capitalizing on the expanding wealth of available data. In contrast, iMD4GC overcomes these challenges, offering a remarkable solution that effortlessly extends to incorporate additional multimodal data. By employing a well-defined data tokenizer, iMD4GC allows for the seamless integration of diverse data sources. This feature is crucial in the current landscape, where an ever-increasing number of modalities are becoming available. The flexible scalability provided by iMD4GC holds immense significance for clinical practice, facilitating precise oncology through AI and multimodal data integration.

This study focuses on four primary modalities: clinical records, pathology images, radiology images, and genomic profiles, which collectively provide crucial information for predicting NACT response and survival prognosis. Nevertheless, it is imperative to acknowledge the existence of additional modalities generated during the medical journey of patients, including endoscopy images and clinical reports. Integrating these additional modalities has the potential to yield valuable insights and enhance the predictive capabilities of our models. Endoscopy images, for instance, capture distinctive visual features that contribute to a comprehensive understanding of the disease, while clinical reports provide expert interpretations of tumor histology. In our future pursuits, we intend to broaden our multimodal learning framework to include these modalities, facilitating a more comprehensive and precise prediction of response and survival prognosis. Through the integration of these diverse data sources, we foresee enhanced accuracy and effectiveness in our predictive models, ultimately benefiting clinical decision-making and patient care. Moving forward, our future endeavors also involve expanding the sample pool and constructing an extensive, large-scale dataset. By doing so, we aim to further increase the capacity of our model and enhance the predictive capabilities of our model. Alongside increasing the scale of the dataset, we will explore the incorporation of foundation models, which possess powerful feature extraction and multimodal information fusion abilities. This integration is expected to significantly improve the performance and robustness of our model in handling multimodal data.

In conclusion, this research presents iMD4GC, a multimodal learning model explicitly designed to address the challenges posed by incomplete multimodal data. It effectively enables precise predictions of NACT response and survival outcomes in GC patients. The iMD4GC outperformed extensive unimodal learning methods, multimodal learning methods, and missing modality methods in terms of response prediction and survival analysis. Furthermore, the inherent interpretability of iMD4GC provides clinicians with valuable insights, empowering them to make informed decisions and enhance the overall quality of patient care. Our future research directions involve the collection of a larger-scale dataset, the inclusion of additional modalities, the introduction of powerful foundation models, and the expansion to other diseases. These endeavors aim to advance the field by increasing the breadth and depth of our understanding, ultimately leading to prediction improvements and better healthcare outcomes.

\section{Methods}
In this section, we first introduce the problem formulation and then present the overall framework of our proposed multimodal learning framework (iMD4GC). Subsequently, we describe the details of each component within the framework, including data tokenization process, structure of unimodal attention layer, and structure of cross-modal interaction layer. Finally, we elaborate on the knowledge distillation employed to enhance the performance of our framework on severely incomplete multimodal data.

\begin{figure*}[tbp]
    \centering
    \includegraphics[width=0.99\linewidth]{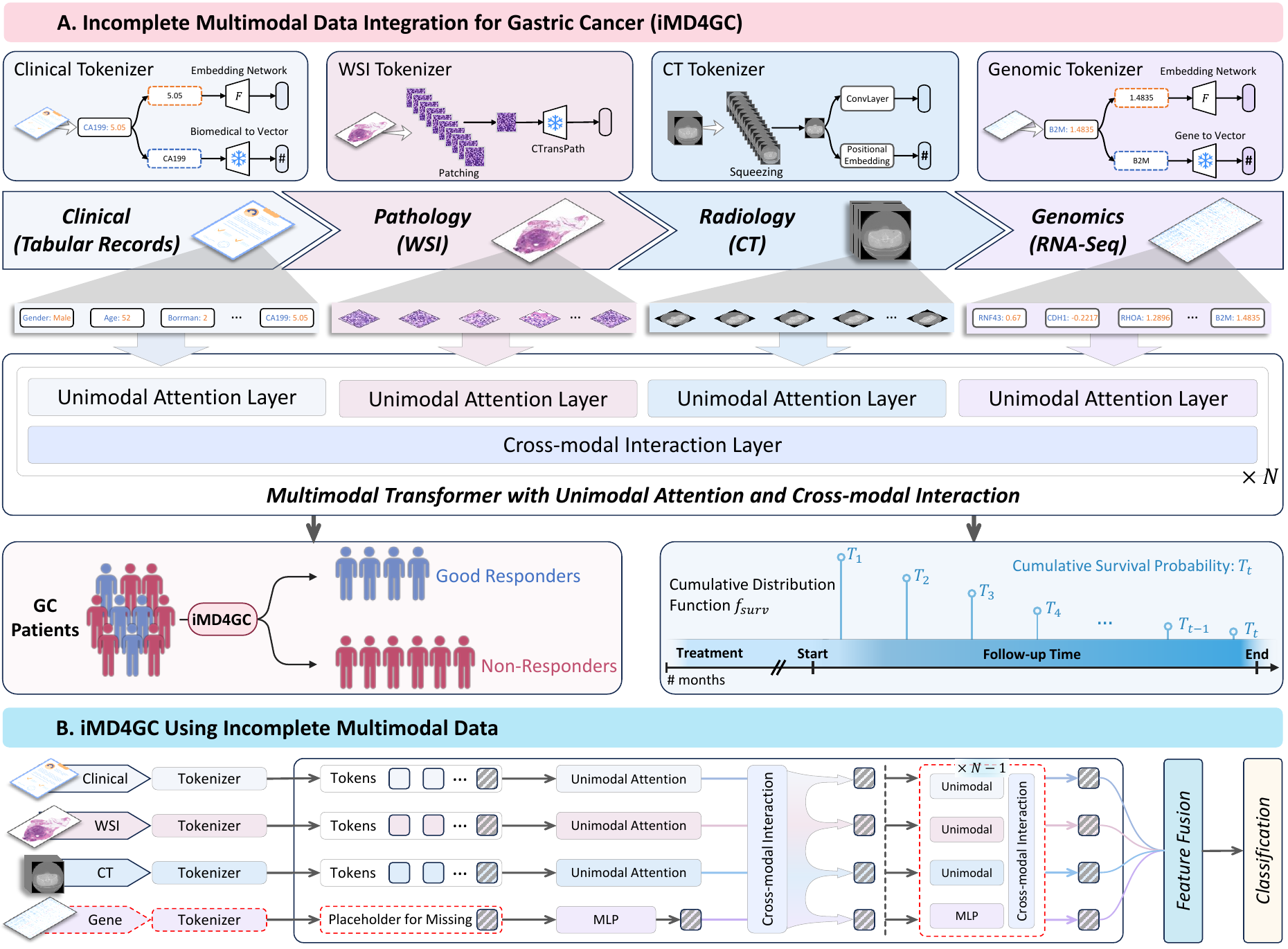}
    \caption{\textbf{Overall framework.} (\textbf{A}) The pipeline of our proposed incomplete multimodal data integration framework (iMD4GC) for NACT response prediction and survival analysis. (\textbf{B}) The pipeline of proposed framework using complete multimodal data. (\textbf{C}) The pipeline of proposed framework using incomplete multimodal data, which explores related information contained in the available modalities and compensates for the missing modalities.}
    \label{fig:framework}
\end{figure*}

\subsection{Problem formulation}
Let $\mathbb{X}=\{\mathcal{X}_1,\mathcal{X}_2,\cdots,\mathcal{X}_N\}$ denote the set of all cases diagnosed with gastric cancer, where $N$ represents the total number of cases. Each individual case can be described by a 4-tuple $\mathcal{X}_i=\{\mathcal{C}_i,\mathcal{R}_i,\mathcal{P}_i,\mathcal{G}_i\}$, in which $\mathcal{C}_i,\mathcal{R}_i,\mathcal{P}_i$, and $\mathcal{G}_i$ correspond to the clinical records, radiology image (CT scans), pathology image (whole slide image), and genomic profiles (RNA-seq), respectively. Let $\mathbb{Y}=\{\mathcal{Y}_1,\mathcal{Y}_2,\cdots,\mathcal{Y}_N\}$ denote the set of all labels, where $\mathcal{Y}_i$ represents the label of $\mathcal{X}_i$. This study encompasses two primary tasks related to gastric cancer: NACT response prediction and survival prediction. In the context of NACT response prediction, the label $\mathcal{Y}$ takes binary values, specifically $0$ to indicate non-responder and $1$ to indicate a good responder. The main objective is to estimate the response probability $\hat{\mathcal{Y}}$. For survival prediction, the label $\mathcal{Y}$ is represented as $\mathcal{Y}=(c,t_{os})$, where $c\in\{0,1\}$ signifies the right uncensored status, and $t_{os}\in\mathbb{R}^+$ represents the overall survival time measured in months. In this case, our primary goal is to estimate the hazard function $f_{hazard}(T=t|T\geq t,\mathcal{X})\in[0, 1]$. The hazard function measures the instantaneous risk or death event occurring at a specific time point $t$. Instead of directly estimating $t_{os}$, survival models output an ordinal risk value obtained by leveraging the cumulative distribution function $f_{surv}(T\geq t,\mathcal{X})=\prod_{u=1}^{t}(1-f_{hazard}(T=t|T\geq t,\mathcal{X}))$.

In practical scenarios, it is common to encounter incomplete modality cases where certain modalities are missing for specific patients. For instance, $\mathcal{X}=\{\mathcal{C},\mathcal{R},\mathcal{P}\}$ indicates the absence of genomic profiles, while $\mathcal{X}=\{\mathcal{C},\mathcal{R}\}$ suggests the lack of both pathology images and genomic profiles. In such cases, the multimodal learning model $\mathcal{F}$ is expected to make accurate predictions based on the available modalities. The multimodal learning model $\mathcal{F}$ should possess the capability to handle both complete and incomplete cases. In this study, we propose a novel multimodal learning framework that effectively integrates all available modalities within $\mathcal{X}$ to estimate the response probability $\hat{\mathcal{Y}}$ and the hazard function $f_{hazard}$ for both complete and incomplete cases.

\subsection{Network architecture}
The overall framework of our proposed multimodal learning model is illustrated in Figure~\ref{fig:framework}A. This model consists of four primary components: data tokenization, unimodal attention layers, cross-modal interaction layers, and multi-modal feature fusion.

\subsubsection{Data tokenization}
It is worth noting that the multimodal data in this work are heterogeneous, encompassing clinical records, radiology images, pathology images, and genomic profiles. To facilitate subsequent modeling tasks, it is necessary to convert these data into a unified format. To this end, we employ different tokenization strategies for different modalities to transform the multimodal data into a set of tokens. Specifically, clinical records $\mathcal{C}$ comprise patient-specific information such as demographics, molecular biomarkers acquired through serological testing, \textit{etc}. These tabular records encompass a diverse array of data types, including discrete, continuous, and collective data. Each clinical record in $\mathcal{C}$ is represented as a 2-tuple $(k,v)$  with $k$ denoting the record's name and $v$ signifying the corresponding value. To explore semantic correlations among these records, we employ BioWordVec~\cite{zhang2019biowordvec} to obtain $d$-dimensional word embeddings for each record, with $d$ representing the embedding dimension. Furthermore, a learnable embedding network is utilized to map the records' values to $d$-dimensional vectors. By integrating word embeddings and value embeddings, clinical records can be tokenized into a set $\mathcal{C}=\{c_1,c_2,\cdots,c_k\}$, with each $c_i\in\mathbb{R}^d$ representing the sum of word embedding and value embedding of the $i$-th record.

The radiology images $\mathcal{R}$ are represented as 3D-CT images, which provides valuable insights into the macroscopic features, morphological characteristics, and tumor texture. However, the stomach only occupies a small portion of the image, with surrounding organs and tissues being irrelevant to prediction tasks. To eliminate extraneous information and focus on the stomach region, we employ TotalSegmentator~\cite{wasserthal2023totalsegmentator} to locate and crop the stomach region from the entire volume. Each slice in the cropped volume is fed into ResNet~\cite{he2016deep} pre-trained on ImageNet to obtain $d$-dimensional image embedding. To maintain spatial correlation among slices, we follow Transformer~\cite{vaswani2017attention} to generate positional embeddings for each slice.  The radiology image can be tokenized into a set $\mathcal{R}=\{r_1,r_2,\cdots,r_m\}$, where each $r_i\in \mathbb{R}^d$ represents the sum of image embedding and positional embedding of the $i$-th slice.

In computational pathology, whole slide images (WSIs) are typically represented as a bag data structure due to their ultra-high resolution. Following this widely used setting, we crop each WSI into a series of non-overlapping $512\times 512$ patches at $40\times$ magnification level. Subsequently, each patch is fed into CTransPath~\cite{wang2022transformer} to obtain $d$-dimensional patch embeddings. Then, pathology image can be tokenized into a set $\mathcal{P}=\{p_1,p_2,\cdots,p_n\}$, with $p_i\in\mathbb{R}^d$ representing the embedding of the $i$-th patch. Similar to clinical records, the genomics profiles can also be regarded as a kind of tabular data, in which each element represents the expression level of a specific gene. To investigate potential co-expression among genes, we adopt Gene2Vec~\cite{du2019gene2vec} to generate $d$-dimensional gene embeddings for each gene. For expression levels, there is a learnable embedding network to map them to $d$-dimensional vectors. By integrating gene embeddings and expression embeddings, genomic profiles can be tokenized into a set $\mathcal{G}=\{g_1,g_2,\cdots,g_l\}$, with each $g_i\in\mathbb{R}^d$ representing the sum of gene embedding and expression embedding of the $i$-th gene.

\begin{figure*}[tbp]
    \centering
    \includegraphics[width=1.0\linewidth]{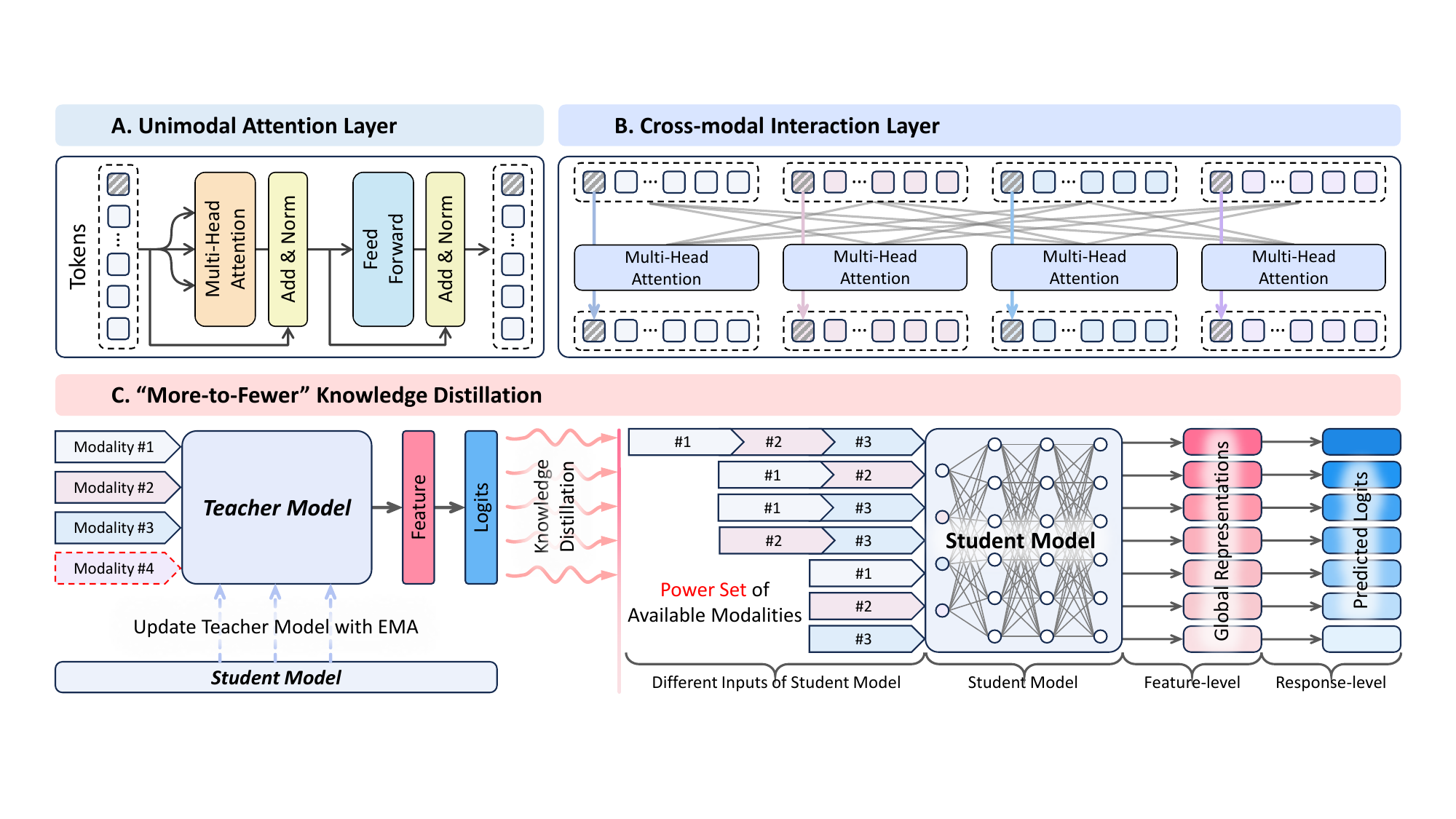}
    \caption{\textbf{Structures of key components and procedure of proposed knowledge distillation.} (\textbf{A}) The structure of unimodal attention layer.  (\textbf{B}) The structure of the cross-modal interaction layer. (\textbf{C}) The proposed ``more-to-fewer'' knowledge distillation transfers knowledge from the teacher model to the student models. The parameters of the teacher model are updated using exponential moving average (EMA), and all student models share the same parameters. The student models take the power set of the available modalities as inputs.}
    \label{fig:KD}
\end{figure*}

\subsubsection{Unimodal attention extracts intra-modal information}
In light of the inherent heterogeneity present in the multimodal data considered in this study, directly fusing all representations from different modalities would lead to irreversible information loss and degradation of performance. Hence, it becomes crucial to identify and extract the most informative and relevant features from each modality. To this end, we leverage unimodal attention layers tailored for each modality, enabling the extraction of discriminative representations from the tokenized data, as shown in Figure~\ref{fig:KD}A. The formulation of the unimodal attention layer can be expressed as follows,
\begin{align}
    (c_*^{(1)}, c_1^{(1)}, c_2^{(1)}, \cdots, c_k^{(1)}) & =\mathcal{F}_{uni}^c(c_*^{(0)}\|\mathcal{C}^{(0)}) \\
    (r_*^{(1)}, r_1^{(1)}, r_2^{(1)}, \cdots, r_k^{(1)}) & =\mathcal{F}_{uni}^r(r_*^{(0)}\|\mathcal{R}^{(0)}) \\
    (p_*^{(1)}, p_1^{(1)}, p_2^{(1)}, \cdots, p_k^{(1)}) & =\mathcal{F}_{uni}^p(p_*^{(0)}\|\mathcal{P}^{(0)}) \\
    (g_*^{(1)}, g_1^{(1)}, g_2^{(1)}, \cdots, g_k^{(1)}) & =\mathcal{F}_{uni}^g(g_*^{(0)}\|\mathcal{G}^{(0)})
\end{align}
where $\mathcal{C}^{(0)},\mathcal{R}^{(0)},\mathcal{P}^{(0)}$, and $\mathcal{G}^{(0)}$ represent the initial tokens of each modality. $c_*^{(0)},r_*^{(0)},p_*^{(0)}$, and $g_*^{(0)}$ denote the learnable class tokens defined for clinical records, radiology images, pathology images, and genomic profiles, respectively. These class tokens serve as placeholders when a particular modality is missing, facilitating subsequent cross-modal information aggregation.

Following the conventional layers in Transformer~\cite{vaswani2017attention}, the unimodal attention layer consists of several primary components: a multi-head self-attention module, a feed-forward network, and layer normalization. This configuration allows the model to effectively capture and process the relationships within each modality. Figure~\ref{fig:framework}B illustrates the structure of the unimodal attention layer within our framework. However, due to the substantial sequence length of pathology and genomic tokens, the native self-attention mechanism becomes computationally expensive and memory-intensive. To address this challenge, we adopt the linear attention mechanism proposed in Nystromformer~\cite{xiong2021nystromformer}, which can effectively reduce the time and space complexity of self-attention while maintaining the overall performance.

\subsubsection{Cross-modal interaction explores potential inter-modal correlations}
The inclusion of unimodal attention layers in our model allows for the extraction of discriminative representations from each modality. However, these representations are inherently limited to their respective modalities and fail to capture the intricate interactions that exist among different modalities. To address this limitation and harness the unique strengths of each modality while integrating information from multiple sources, we propose the incorporation of cross-modal interaction layers in our framework. These interaction layers follow the unimodal attention layers and enable the exploration of inter-modal interactions and the aggregation of inter-modal information. To ensure the integrity of the information and mitigate any contamination arising from the heterogeneity of multimodal data, we leverage the class tokens $\{c_*^{(1)},r_*^{(1)},p_*^{(1)}, g_*^{(1)}\}$ as the query tokens, as shown in Figure~\ref{fig:KD}B. These class tokens serve as the information bridge for exploring inter-modal interactions and aggregating inter-modal information. The formulation of the cross-modal interaction layer can be expressed as follows,
\begin{align}
    \hat{c}_*^{(1)} & =\mathcal{F}_{cross}^c(c_*^{(1)}\|\mathcal{R}^{(1)}\|\mathcal{P}^{(1)}\|\mathcal{G}^{(1)}) \\
    \hat{r}_*^{(1)} & =\mathcal{F}_{cross}^r(r_*^{(1)}\|\mathcal{C}^{(1)}\|\mathcal{P}^{(1)}\|\mathcal{G}^{(1)}) \\
    \hat{p}_*^{(1)} & =\mathcal{F}_{cross}^p(p_*^{(1)}\|\mathcal{C}^{(1)}\|\mathcal{R}^{(1)}\|\mathcal{G}^{(1)}) \\
    \hat{g}_*^{(1)} & =\mathcal{F}_{cross}^g(g_*^{(1)}\|\mathcal{C}^{(1)}\|\mathcal{R}^{(1)}\|\mathcal{P}^{(1)})
\end{align}
where $\mathcal{C}^{(1)},\mathcal{R}^{(1)},\mathcal{P}^{(1)}$, and $\mathcal{G}^{(1)}$ represent the outputs of the previous unimodal attention layers. It is worth noting that, we only need to calculate the attention between the class token and the other tokens, which significantly reduces the computational complexity compared to the unimodal attention layer. Hence, we can directly employ the naive multi-head attention mechanism. Our proposed model alternates between stacking unimodal attention and cross-modal interaction layers, which enables the extraction of informative representations from individual modalities while effectively capturing the complex interactions that exist among different modalities. By incorporating both types of attention layers, our model can generate comprehensive representations that encompass the unique characteristics of each modality and the intermodal relationships between them.

\subsubsection{Cross-modal interaction provides inter-modal complementary information}
When certain modalities are unavailable, there are two primary strategies for handling incomplete multimodal data: 1) discarding the missing modalities and only utilizing the available modalities, and 2) aggregating complementary information from the available modalities to compensate for the missing modalities. The first strategy may seem straightforward, involving the removal of corresponding attention layers, but it inevitably leads to a loss of valuable information and a decline in performance. In contrast, the second strategy holds promise for achieving superior performance. In our framework, we embrace the second strategy and leverage the power of cross-modal interaction layers to effectively compensate for the missing modalities. As depicted in Figure~\ref{fig:framework}B, the class token serves as a placeholder when a specific modality is missing. In such cases, the corresponding unimodal attention layer is degraded to a linear layer. Consequently, the placeholder is seamlessly integrated into the cross-modal interaction layer, enabling the capture of complementary information from the available modalities. This process empowers our model to compensate for the missing modalities, thereby preserving the accuracy and reliability of its predictions. By leveraging the available information and exploiting the intricate relationships among different modalities, our framework could surmount the limitations imposed by incomplete multimodal data, ultimately enhancing the robustness and efficacy of our predictions.

\subsection{``More-to-fewer'' knowledge distillation}
To further enhance the performance of our framework when dealing with severely incomplete multimodal data, we introduce a ``more-to-fewer'' knowledge distillation, which aims to distill knowledge from all available modalities to the reduced sets (power set of $\mathcal{X}$). The strategy consists of three steps: 1) training a teacher model $\mathcal{F}_t$ using all available modalities, 2) constructing a student model $\mathcal{F}_s$ that only utilizes a subset of modalities as input, and 3) transferring knowledge from the teacher model to the student model. Figure~\ref{fig:KD}C illustrates the overall framework of this strategy. Within this work, two types of knowledge distillation are employed: feature-level distillation and response-level distillation (logits-level). Feature-level distillation ensures that the student model generates representations similar to those of the teacher model, while response-level distillation focuses on ensuring that the student model produces predictions comparable to those of the teacher model. The knowledge distillation process can be formulated as follows:
\begin{align}
    \mathcal{L}_{fea} & = \sum_{S\in \mathcal{P}^*(\mathcal{X})}\mathcal{D}_{KL}(\mathcal{F}_\mathcal{X},\mathcal{F}_\mathcal{S})             \\
    \mathcal{L}_{res} & = \sum_{S\in \mathcal{P}^*(\mathcal{X})}\mathcal{D}_{KL}(\hat{\mathcal{Y}}_\mathcal{X},\hat{\mathcal{Y}}_\mathcal{S})
\end{align}
where $\mathcal{P}^*(\mathcal{X})$ represents the power set of $\mathcal{X}$ excluding the empty set. $\mathcal{D}_{KL}$ denotes the Kullback-Leibler divergence loss function. $\mathcal{F}_\mathcal{X}$ and $\mathcal{F}_\mathcal{S}$ are the representations generated by the teacher model and the student model, respectively. $\hat{\mathcal{Y}}_\mathcal{X}$ and $\hat{\mathcal{Y}}_\mathcal{S}$ denote the predictions made by the teacher model and the student model, respectively. To mitigate heavy computation requirements and reduce the risk of overfitting, we employ exponential moving average (EMA)~\cite{tarvainen2017mean} to update the parameters of the teacher model. Additionally, the student model are updated while being constrained by the classification loss. Incorporating classification loss with knowledge distillation loss, the student model can learn from the teacher model and achieve comparable performance even in scenarios where more modalities are missing. This strategy enhances the effectiveness of our framework in handling severely incomplete multimodal data.

\backmatter

\bmhead{Supplementary information}
We have two accompanying supplementary files:
\begin{itemize}
    \item Appendix A: Datasets
          \begin{itemize}
              \item A1: GastricRes
              \item A2: GastricSur
              \item A3: TCGA-STAD
          \end{itemize}
    \item Appendix B: Method details
          \begin{itemize}
              \item B1: Loss functions
              \item B2: Implementation details
              \item B3: Ablation study
          \end{itemize}
\end{itemize}

\bmhead{Acknowledgments}
This work was supported by National Natural Science Foundation of China (No. 62202403, 82001986, and 82360345), Hong Kong Innovation and Technology Fund (No. PRP/034/22FX), Shenzhen Science and Technology Innovation Committee Funding (Project No. SGDX20210823103201011), the Research Grants Council of the Hong Kong Special Administrative Region, China (Project No. R6003-22 and C4024-22GF).

\bmhead{Data availability}
There are three datasets involved in this study, including GastricRes for NACT response prediction, GastricSur for survival analysis, and TCGA-STAD for survival analysis. The first two datasets are not publicly released due to restrictions by privacy concern, but they are available from the corresponding author on reasonable request. The TCGA-STAD dataset is publicly available at \url{https://portal.gdc.cancer.gov/}. The formatted data of TCGA-STAD is available at \href{https://hkustconnect-my.sharepoint.com/:f:/g/personal/fzhouaf_connect_ust_hk/ElYrNphQdy9GrPvkalfqIr4BMx7s8EQk3EsLIEcEizFGhA?e=Cbr1fn}{OneDrive}, which can be directly used to reproduce the experimental results in this study.

\bmhead{Code availability}
All code was implemented in Python using PyTorch as the primary deep learning package. All code and scripts to reproduce the experiments of this study are available at \url{https://github.com/FT-ZHOU-ZZZ/iMD4GC}.

\bmhead{Declaration of interests}
The authors have no conflicts of interest to declare.










\begin{appendices}
    \section{Datasets}\label{secA1}
To evaluate the performance of our framework, we collected three multimodal datasets, including GastricRes, GastricSur, and TCGA-STAD. In this section, we introduce the detail of each dataset.

\begin{figure*}[tbp]
    \centering
    \includegraphics[width=1.0\linewidth]{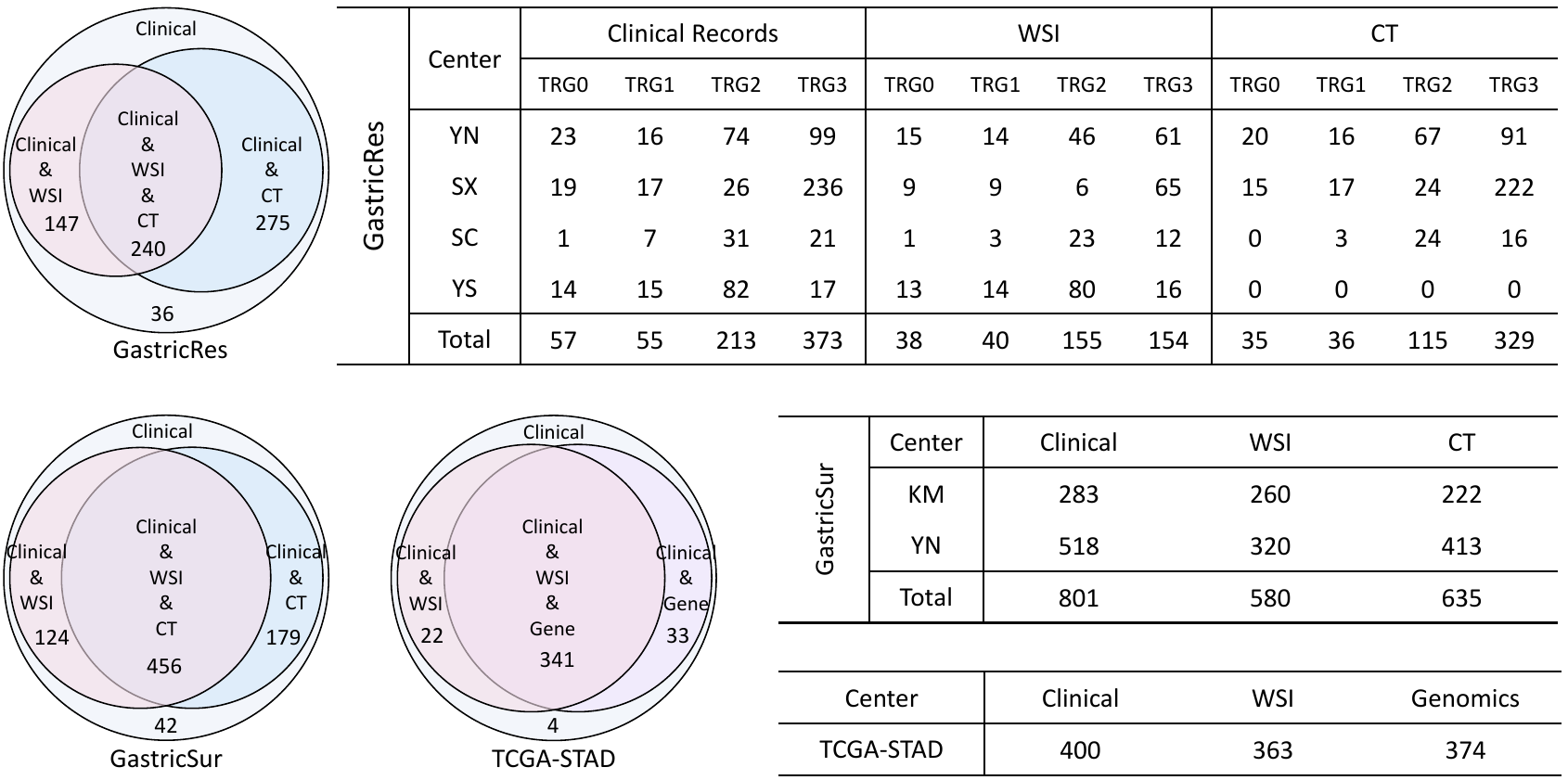}
    \caption{\textbf{Details of datasets used in this study.} There are three datasets involved in this study: GastricRes dataset for response prediction, GastricSur dataset for survival analysis, and TCGA-STAD dataset for survival analysis. GastricRes comprises information from 698 patients diagnosed with gastric cancer who underwent NACT treatment, including three modalities: clinical records, WSI, and CT. GastricSur consists of data from 801 patients diagnosed with gastric cancer who underwent surgical resection, including three modalities: clinical records, WSI, and CT. TCGA-STAD contains data from 400 patients diagnosed with gastric cancer, including three modalities: clinical records, WSI, and genomic profiles.}
    \label{fig:dataset}
\end{figure*}

\subsection{GastricRes}
This dataset was collected from four prominent medical hospitals in China: Yunnan Cancer Hospital (Kunming, China), Shanxi Cancer Hospital (Taiyuan, China), Sichuan Cancer Hospital (Chengdu, China), and the Sixth Affiliated Hospital of Sun Yat-sen University (Guangzhou, China). This dataset encompasses comprehensive information from 698 patients who were diagnosed with gastric cancer and underwent Neoadjuvant Chemotherapy (NACT) treatment. More details can be found in Figure~\ref{fig:dataset}. The dataset consists of three modalities: clinical records, whole slide images (WSI), and computed tomography images (CT). Tumor Regression Grade (TRG) is a widely used grading system employed to evaluate the extent of tumor regression and the response to NACT in patients with gastric cancer. TRG0 represents the absence of residual tumor cells observed under microscopic examination, indicating a pathologically complete response. TRG1 indicates the presence of only single cells or small clusters of cancer cells visible under the microscope. TRG2 denotes the presence of residual cancer cells surrounded by fibrosis. TRG3 indicates significant fibrosis predominating over cancer cells. In this study, we define TRG(0-2) as a good responder, indicating favorable treatment response, while TRG3 is classified as a non-responder, indicating limited response to therapy. There are 325 good responders and 373 non-responders in the GastricRes dataset.

\subsection{GastricSur}
This dataset was collected from two prominent medical hospitals in China: the First Affiliated Hospital of Kunming Medical University (Kunming, China) and Yunnan Cancer Hospital. It comprises comprehensive data from a cohort of 801 patients who were diagnosed with gastric cancer and subsequently underwent surgical resection. Detailed information regarding the dataset can be found in Figure~\ref{fig:dataset}. Similar to the GastricRes dataset, this collection encompasses three distinct modalities: clinical records, whole slide images (WSI), and computed tomography scans (CT). Throughout the follow-up period, there are 286 patients who died.

\subsection{TCGA-STAD}
This dataset was obtained from The Cancer Genome Atlas (TCGA) database. It contains data from 400 patients diagnosed with gastric cancer. Different from GastricRes and GastricSur datasets, the modalities involved in this dataset are: clinical records, WSI, and RNA-seq. The clinical records are downloaded from the \href{https://linkedomics.org/data_download/TCGA-STAD/}{LinkedOmics}. The WSI are downloaded from the \href{https://portal.gdc.cancer.gov/}{GDC Data Portal}. The RNA-seq data is downloaded from the \href{https://www.cbioportal.org/}{cBioPortal}. Specifically, all patients in this dataset have clinical records, while 363 patients have WSI and 374 patients have RNA-seq.


\section{Method details}\label{secA2}
\subsection{Loss functions}
There are two tasks in this study: NACT response prediction and survival prediction. The former is a classification task, while the latter is a regression task. We employ different loss functions for these two tasks. In addition to the loss function defined for specific tasks, there is also a loss function defined for knowledge distillation.

\textbf{NACT response prediction.} We employ the cross-entropy function to constrain the NACT response prediction task. The cross-entropy loss function is defined as follows:
\begin{align}
    D_{\text{ce}}(\mathcal{Y}, \hat{\mathcal{Y}}) = -\sum_{i=1}^{C} \mathcal{Y}_i \log(\hat{\mathcal{Y}}_i)
\end{align}
where $\mathcal{Y}$ is the ground truth label, $\hat{\mathcal{Y}}$ is the predicted label, and $\log$ is the natural logarithm. There are two types of knowledge distillation: feature-level distillation and response-level distillation. Feature-level distillation ensures that the student models generate representations similar to those of the teacher model, while response-level distillation focuses on ensuring that the student models produce predictions comparable to those of the teacher model. We employ the Kullback-Leibler divergence to constrain the knowledge distillation process. For feature-level distillation, the Kullback-Leibler divergence is defined as follows:
\begin{align}
    D_{\text{KL}}(\mathcal{F}_t \| \mathcal{F}_s) = \sum_{i=1}^{K} (\sigma(\frac{\mathcal{F}_t}{T})_i \log(\frac{\sigma(\frac{\mathcal{F}_t}{T})_i}{\sigma(\frac{\mathcal{F}_s}{T})_i}))
\end{align}
where $\mathcal{F}_t$ and $\mathcal{F}_s$ are the feature representations of the teacher model and student model, respectively. $T$ is the temperature parameter. $\sigma$ is the softmax function. $K$ is the dimension of features. For response-level distillation, the Kullback-Leibler divergence is defined as follows:
\begin{align}
    D_{\text{KL}}(\mathcal{Y}_t \| \mathcal{Y}_s) = \sum_{i=1}^{C} (\sigma(\frac{\mathcal{Y}_t}{T})_i \log(\frac{\sigma(\frac{\mathcal{Y}_t}{T})_i}{\sigma(\frac{\mathcal{Y}_s}{T})_i}))
\end{align}
where $\mathcal{Y}_t$ and $\mathcal{Y}_s$ are the predicted labels of the teacher model and student model, respectively. $T$ is the temperature parameter. $\sigma$ is the softmax function. $C$ is the number of classes. In the proposed ``more-to-less'' knowledge distillation, the teacher model is trained using all available modalities, while the student models are trained using a subset of modalities. The knowledge distillation process can be formulated as follows:
\begin{align}
    \mathcal{L}_{cls} & = \sum_{S\in \mathcal{P}^*(\mathcal{X})}D_{\text{ce}}(\mathcal{Y}, \hat{\mathcal{Y}}_\mathcal{S})                       \\
    \mathcal{L}_{fea} & = \sum_{S\in \mathcal{P}^*(\mathcal{X})}\mathcal{D}_{KL}(\mathcal{F}_\mathcal{X}\| \mathcal{F}_\mathcal{S})             \\
    \mathcal{L}_{res} & = \sum_{S\in \mathcal{P}^*(\mathcal{X})}\mathcal{D}_{KL}(\hat{\mathcal{Y}}_\mathcal{X}\| \hat{\mathcal{Y}}_\mathcal{S})
\end{align}
where $\mathcal{X}$ is the set of all available modalities. $\mathcal{P}^*(\mathcal{X})$ represents the power set of $\mathcal{X}$ excluding the empty set. $\mathcal{F}_\mathcal{X}$ and $\mathcal{F}_\mathcal{S}$ are the representations generated by the teacher model and the student model, respectively. $\hat{\mathcal{Y}}_\mathcal{S}$ denotes the prediction made by the student model. The total loss function is defined as follows:
\begin{align}
    \mathcal{L} = \mathcal{L}_{cls} + \alpha\mathcal{L}_{fea} + \beta\mathcal{L}_{res}
\end{align}
where $\alpha$ and $\beta$ are the hyperparameters.

\textbf{Survival prediction.} Due to the huge size of WSI, the model cannot be optimized with mini-batch manner. The alternative optimization strategy is to consider discrete time intervals and model each interval using an independent output. We leverage NLL (negative log-likelihood) survival loss~\cite{chen2021multimodal} as the loss function of the survival prediction part. The loss functions for knowledge distillation are the same as those defined for the NACT response prediction task. The total loss function is defined as follows:
\begin{align}
    \mathcal{L}_{sur} = \sum_{S\in \mathcal{P}^*(\mathcal{X})}\mathcal{L}_{NLL}(T_S, l, c) \\
    \mathcal{L}       = \mathcal{L}_{sur} + \alpha\mathcal{L}_{fea} + \beta\mathcal{L}_{res}
\end{align}
where $\mathcal{L}_{NLL}$ is the NLL survival loss. $T_S=\{T_1, T_2,\cdots,T_i\}$ is the predicted hazard value at each discrete time interval. $l$ is the discrete label. $c$ is the censoring status. $\alpha$ and $\beta$ are the hyperparameters.

\begin{table*}
    \begin{center}
        \caption{\textbf{Impacts of feature fusoon strategies.} This table presents the results of different feature fusion strategies on the GastricRes, GastricSur, and TCGA-STAD datasets. The best results are highlighted in \textbf{bold}, while the second-best results are \underline{underlined}.}
        \begin{tabular}{lccccc}
            \toprule
            {\multirow{2}{*}{Method}} & \multicolumn{3}{c}{GastricRes} & GastricSur                     & TCGA-STAD                                                                                        \\\cmidrule(lr){2-4}\cmidrule(lr){5-5}\cmidrule(lr){6-6}
                                      & AUC                            & ACC                            & F1-Score                       & C-Index                        & C-Index                        \\
            \midrule
            ConcatWithLinear          & \textbf{0.802$_{\pm0.050}$}    & \textbf{0.758$_{\pm0.043}$}    & \textbf{0.753$_{\pm0.027}$}    & \textbf{0.714$_{\pm0.008}$}    & \textbf{0.661$_{\pm0.058}$}    \\
            Multiplicative            & 0.760$_{\pm0.051}$             & \underline{0.699$_{\pm0.067}$} & \underline{0.694$_{\pm0.066}$} & 0.540$_{\pm0.031}$             & 0.589$_{\pm0.064}$             \\
            TensorFusion              & 0.759$_{\pm0.029}$             & 0.588$_{\pm0.051}$             & 0.462$_{\pm0.129}$             & 0.691$_{\pm0.018}$             & \underline{0.656$_{\pm0.048}$} \\
            LowRankTensorFusion       & 0.790$_{\pm0.055}$             & 0.669$_{\pm0.099}$             & 0.623$_{\pm0.145}$             & \underline{0.705$_{\pm0.015}$} & 0.653$_{\pm0.043}$             \\
            EarlyFusionTransformer    & \underline{0.797$_{\pm0.060}$} & 0.688$_{\pm0.068}$             & 0.671$_{\pm0.075}$             & 0.703$_{\pm0.010}$             & 0.643$_{\pm0.054}$             \\
            LateFusionTransformer     & 0.772$_{\pm0.059}$             & 0.623$_{\pm0.079}$             & 0.522$_{\pm0.144}$             & 0.565$_{\pm0.023}$             & 0.596$_{\pm0.036}$             \\
            \bottomrule
        \end{tabular}
    \end{center}
    \label{tab:fusion}
\end{table*}

\begin{figure*}[tbp]
    \centering
    \includegraphics[width=1.0\linewidth]{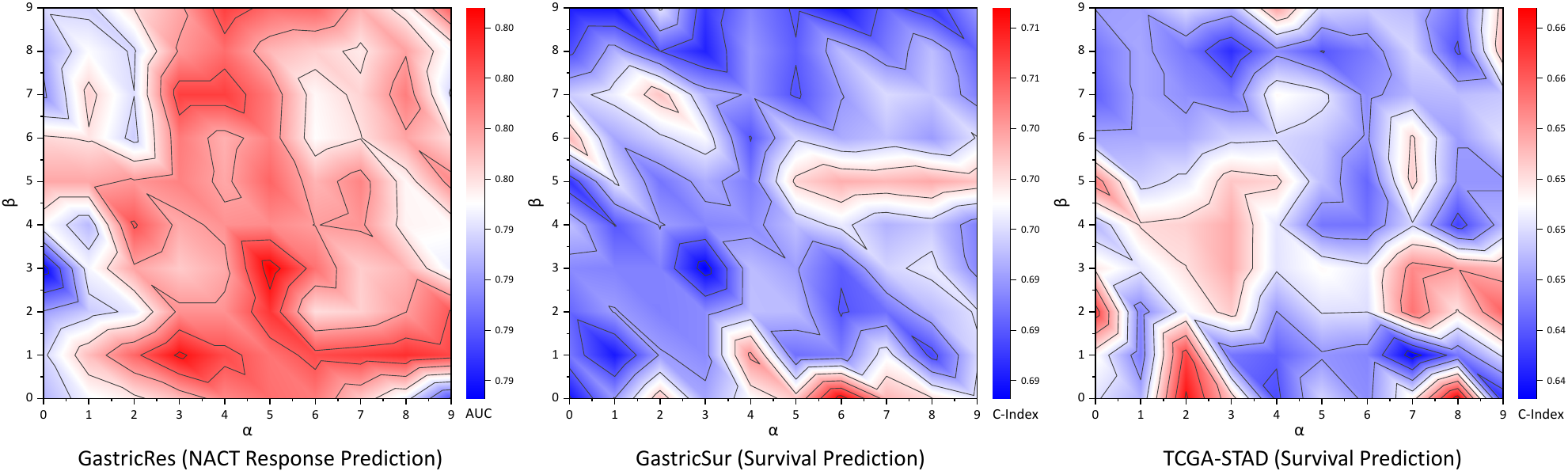}
    \caption{\textbf{Impacts of hyperparameters.} These heatmap figures present the results of different hyperparameters on the GastricRes, GastricSur, and TCGA-STAD datasets. The x-axis represents $\alpha$ for feature-level distillation, while the y-axis represents  $\alpha$ for response-level distillation.}
    \label{fig:parameter}
\end{figure*}

\subsection{Implementation details}
The proposed framework was implemented using PyTorch 1.12.0 and Python 3.9.12. The complete source code can be accessed and reviewed at the following URL: \url{https://github.com/FT-ZHOU-ZZZ/iMD4GC}. To ensure reproducibility and minimize the impact of randomness on experimental results, the random seeds of both PyTorch and NumPy were set to a fixed value of 1, providing a consistent basis for comparison across experiments. In the clinical part, each record was fed into a 200-dimensional fully connected layer to obtain 200-dimension embedding. Similarly, in the pathology part, CTransPath~\cite{wang2022transformer} was employed to extract 768-dimensional patch features, which were then fed into a series of fully connected layers (768-256-200) to obtain compact and informative 200-dimensional embeddings. In the radiology part, the feature maps of penultimate layer from the ResNet-50 architecture were utilized. These feature maps were processed through a global average pooling layer to generate a concise 256-dimension feature vector. Subsequently, this feature vector was fed into a sequence of fully connected layers (256-256-200) to yield comprehensive and representative 200-dimensional embeddings. Regarding the genetic part, the expression level of each gene was fed into a 200-dimensional fully connected layer to obtain 200-dimension embedding.  The dimension of all tokens was set to 200, which is equal to the dimension of gene embedding and expression embedding. It is widely recognized that augmenting the number of layers and parameters within a model generally amplifies its capacity, leading to improved performance. However, the availability of training samples in our dataset is restricted, thereby requiring a delicate balance between the number of network layers and the available training data to mitigate the risk of overfitting. In this study, we set the number of network layers to 2 for all datasets.

All methods were optimized with Adam optimizer. The initial learning rate is set to 0.0002. To further enhance the training process, a CosineAnnealingLR~\cite{loshchilov2016sgdr} scheduler was employed to dynamically adjust the learning rate, promoting convergence and preventing overfitting. Due to the different number of tokens, this framework cannot be trained with mini-batch manner. Therefore, the batch size is set to 1. The maximum number of epochs is set to 30. The weight decay is set to 0.00001. The temperature parameter in knowledge distillation is set to 4. More details can be found in the scripts of \url{https://github.com/FT-ZHOU-ZZZ/iMD4GC}. All methods involved in this study were trained and validated on a high-performance workstation with 8 NVIDIA RTX 3090 GPUs.

\subsection{Ablation study}
\subsubsection{Impacts of feature fusion}
As Ma \textit{et al.}~\cite{ma2022multimodal} pointed out, different fusion strategies do affect the robustness of Transformer models. Surprisingly, the optimal fusion strategy is dataset-dependent; there does not exist a universal strategy that works in general cases. To evaluate the impacts of feature fusion, we conduct a series of experiments to investigate the performance of different fusion strategies, including concatenation, multiplicative, tensor fusion~\cite{zadeh2017tensor}, and low-rank tensor fusion~\cite{liu2018efficient}, early fusion with transformer~\cite{vaswani2017attention}, and late fusion with transformer. The results are shown in Table~\ref{tab:fusion}. It can be observed that the concatenation achieved the best performance across all datasets. The low-rank tensor fusion and early fusion with transformer also achieved promising results and exhibited potential for further exploration.

\subsubsection{Impacts of hyperparameters}
There are two hyperparameters in our proposed framework: $\alpha$ and $\beta$. $\alpha$ is the weight of feature-level distillation, while $\beta$ is the weight of response-level distillation. It is crucial to carefully select and balance these hyperparameters to enhance the model's performance based on the specific tasks and datasets. To investigate the impacts of these hyperparameters, we conduct a series of experiments to evaluate the performance of different hyperparameters. The results are shown in Figure~\ref{fig:parameter}. It can be observed that $\alpha$ and $\beta$ have a significant impact on the performance of the proposed framework. The optimal hyperparameters are dataset-dependent. For the GastricRes dataset, the optimal hyperparameters are $\alpha=5.0$ and $\beta=3.0$. For the GastricSur dataset, the optimal hyperparameters are $\alpha=6.0$ and $\beta=0.0$. For the TCGA-STAD dataset, the optimal hyperparameters are $\alpha=8.0$ and $\beta=0.0$. That is, the feature-level distillation and response-level distillation are both important for the classification task, \textit{i.e.}, the NACT response prediction task in the GastricRes dataset, while the response-level distillation is more important for regression task, \textit{i.e.}, the survival analysis task in the GastricSur and TCGA-STAD datasets.
\end{appendices}


\bibliographystyle{unsrt}
\bibliography{sn-bibliography}

\end{document}